\documentclass[preprints,article,accept,moreauthors,pdftex,10pt,a4paper]{Definitions/mdpi}
\usepackage{amssymb}
\usepackage{subcaption}
\usepackage{soul}
\newcommand{\bra}[1]{\left< #1 \right|}
\newcommand{\ket}[1]{\left| #1 \right>}
\firstpage{1} 
\pubvolume{xx}
\issuenum{1}
\articlenumber{5}
\pubyear{2020}
\copyrightyear{2020}
\history{Received: date; Accepted: date; Published: date}

\Title{Superfluid neutron matter with a twist}

\Author{Georgios Palkanoglou $^{1}$\orcidA{} and Alexandros Gezerlis $^{1}$\orcidB{}}

\AuthorNames{Georgios Palkanoglou, Alexandros Gezerlis}

\address{%
$^{1}$ \quad Department of Physics, University of Guelph, Guelph, ON N1G 2W1, Canada}

\abstract{
Superfluid neutron matter is a key ingredient in the composition of neutron stars. The physics of the inner crust is largely dependent on that of its $S$-wave neutron superfluid which has made its presence known through pulsar glitches and modifications on neutron star cooling. Moreover, with recent gravitational-wave observations of neutron star mergers, the need for an equation of state for the matter of these compact stars is further accentuated and a model-independent treatment of neutron superfluidity is important. \textit{Ab initio} techniques developed for finite systems can be guided to perform extrapolations to the thermodynamic limit and attain this model-independent extraction of various quantities of infinite superfluid neutron matter. To inform such an extrapolation scheme, we performed calculations of the neutron $^1S_0$ pairing gap using the model-independent odd-even staggering in the context of the particle-conserving, projected BCS theory under twisted boundary conditions. While the practice of twisted boundary conditions is standard in solid state physics and has been used repeatedly in the past to reduce finite-size effects, this is the first time it is employed in the context of pairing. We find that a twist-averaging approach results in a substantial reduction of the finite-size effects, bringing systems with $N\gtrapprox 50$ within a $2\%$ error margin from the infinite system. This can significantly reduce extrapolation-related errors in the extraction of superfluid neutron matter quantities.
}


\begin{document}


\section{Introduction}
Historically, the idea of neutron stars (NSs), i.e., ultra-compact objects where ``the atomic nuclei come in close contact, forming a gigantic nucleus", was first proposed by L.~Landau~\cite{Landau} shortly before~\cite{Yakolev:2013} the discovery of the neutron~\cite{Chadwick}. This idea was further explored by W.~Baade and F.~Zwicky~\cite{Zwicky:1, Zwicky:2} in two seminal publications where they identified the birth of NSs with supernova explosions, a term also coined therein. These theoretical ideas were substantiated by the observation of the first pulsar by J.~Bell~\cite{Bell} which, interpreted by T.~Gold~\cite{Gold} shortly after, marked the first observation of a NS. While the interest in compact stars only grew after that, a parallel thread involved nuclear superfluidity which was first proposed by A.~Bohr, B.~R.~Mottelson, and D.~Pines~\cite{Bohr:1958} a year after the microscopic theory of superfluidity was introduced (see Sec.~\ref{sec:BCS}) and a year before A.~B.~Migdal~\cite{Migdal:1959} remarked that similar mechanisms could take place in the interior of NSs. Today, the presence of supefluidity in NSs is a widely accepted fact~(for a more detailed account on the history of NSs see Refs.~\cite{Yakolev:2013}~\&~\cite{Book:Haensel}, as well as other contributions to this special issue \cite{Durel:2020, Shelley:2020, Wei:2020}).  

While the idea of compact stars was first discussed during the beginning of modern nuclear theory, the details of the composition of these objects came with later advancements in the physics of nuclei and the further understanding of the nuclear interaction~\cite{Book:Haensel}. The current understanding of NSs separates the compact object in layers of different density which also correspond to regions of different physical phenomena. The outer crust of a NS is composed of a lattice of nuclei bathed in a sea of electrons and neutrons at densities $\rho \approx10^6 ~\textrm{g}\,\textrm{cm}^{-3}$. Starting from $^{56}\textrm{Fe}$, as one goes deeper the neutron fraction in the nuclei of the lattice increases as higher density imposes an increasing number of neutrons in each nucleus. Pairing correlations for the neutrons are already present at this point only within the neutron-rich nuclei. The free neutrons permeating the lattice create a state of matter similar to terrestrial binary alloys while also providing an effective attraction between nuclei resulting in clumps similar to those found in metallic alloys~\cite{Kobyakov:2014}, but also see Ref.~\cite{Kobyakov:2016}. Reaching the innermost region of the outer crust, the composition of matter becomes uncertain largely due to the uncertainties in measurements of nuclear masses or the lack of relevant data~\cite{Blascke:2018}. The inner crust of the NS starts at densities where the neutrons start dripping out of the nuclei ($\rho \approx 4\times 10^{11}~\textrm{g}\,\textrm{cm}^{-3}$) resulting in a state of neutron-rich nuclei in a background of ultrarelativistic electrons and a dilute fluid of neutrons~\cite{Chamel:2015}. The inner crust extends to densities up to half the nuclear saturation density, $\rho_0=2.8\times 10^{14}~\textrm{g}\,\textrm{cm}^{-3}$. At these densities the $^1S_0$ channel of the neutron-neutron (NN) interaction becomes attractive~\cite{Stoks:1993} ensuring superfluidity for the neutron fluid outside of the neutron-rich nuclei as well. With the protons mainly confined in the nuclei, in this depth, and far from the neutron-drip transition~\cite{Pastore:2013}, one can approximate the neutron superfluid as pure neutron matter (NM) coupled to the crustal phonons or band structure~\cite{Martin:2014, Martin:2016, Sedrakian:2019}, as far as static properties are concerned~\cite{Chamel:2010} (more generally see Refs.~\cite{Chamel:2013, Watanabe:2017, Chamel:2017, Durel:2018, Inakura:2017, Inakura:2019}). At this point superfluidity exists both inside~\cite{Pastore:2013} and outside the neutron-rich nuclei but this boundary becomes increasingly vague as the density of the neutron fluid increases, i.e., as one reaches the bottom of the inner crust where the deformation of neutron-rich nuclei marks the onset of the nuclear pasta. These extended clusters of neutrons and protons can be shown to be the result of a balance between Coulomb and surface effects, in the context of a liquid drop model~\cite{Baym:1971}. The first order transition leads to an extended pasta phase beyond which lies the outer core of the star at a density close to nuclear saturation. Here the neutron density, while ensuring an $S$-wave repulsion between the neutrons, allows for attractions through other channels of the NN interaction creating $PF$-superfluidity while the proton density reaches that of the neutrons in the inner crust turning the proton fluid into an $S$-superconductor. With densities ranging from $2\rho_0$ to $10\rho_0$, the consistency of the star in the core is largely uncertain. A popular conjecture is the appearance of hyperons such as $\Sigma^{\pm}$, $\Lambda$, and $\Xi^{\pm}$, when the Fermi energy of the neutron and electron gas surpasses their corresponding rest masses. This can also lead to hyperon superfluidity if their interaction is attractive. Another popular conjecture involves quark deconfinement which, due to the uncertainty in the density marking its onset, might come before or after the formation of hyperons. As the quark degrees of freedom start becoming important one might also expect the appearance of quark-gluon superconductivity~\cite{Alford:2008}. Finally, Bose-Einstein condensation of pions and kaons has also been theorized to exist in the core~\cite{Takatsuka:1993, Pethick:arxiv}. For a more detailed review on the current consensus on the composition of NS see Ref.~\cite{Sedrakian:2019}~\&~\cite{Blascke:2018}.

Nuclear superfluids (and superconductors) created by pairing of nucleons in various channels, permeate most layers of a NS and are responsible for a variety of astrophysical phenomena, such as impacts on the cooling of the star~\cite{Yakovlev:2004, Page:2011, Page:2012} and the anomalous glitches observed in the rotation of pulsars~\cite{Haskell:2015}. The physics of the superfluids found in a NS can also affect its seismic properties and, hence, the neutrino and gravitational radiation emitted from it~\cite{Andersson:2021}. Furthermore, pairing has direct consequences on the Equation of State (EoS) of NM \cite{Gandolfi:2008} which in turn determines the mass-radius relation of NSs and their maximum mass. Finally, the radiation from various superfluid phases in NSs can be used for constraints on the coupling of exotic particles, proposed as extensions to the Standard Model (e.g., axions~\cite{Sedrakian:2016}), with Standard Model matter. Thus, the correct description of superfluid NM is an important step in the understanding of the physics of NSs and their connection to the cosmos.

Many models have been proposed for a concise description of superfluid NM generating a polyphony of results and a landscape of pairing gaps~\cite{Sedrakian:2019, Dean:2002}. This only underlines the need for a model-independent extraction of the properties of superfluid NM. Promising candidates for such an extraction are \textit{ab initio} approaches which calculate quantum many-body properties by attacking the problem from first principles. With a sizable part of these being techniques developed for tackling finite systems, one is faced with the task of creating a well-informed extrapolation scheme able to map the properties of the finite system to the infinite one, where NS matter lies. The rest of this paper is organized as follows. In section~\ref{subsec:NM} we present a brief overview of superfluidity in NM, in section~\ref{sec:BCS} we discuss the fundamentals of the BCS theory of superconductivity and its particle-conserving variation PBCS, and how these can form an extrapolation scheme to the TL for \textit{ab initio} approaches. Finally, in section~\ref{sec:FSE},  we present a brief overview of the finite-size effects (FSE) in superfluid NM and we apply techniques of manipulating the periodic boundary conditions (PBC) of a finite system to decrease these FSE and further improve the extrapolation schemes already in use for NM. We present results demonstrating that approaches such as twisting the boundary conditions (BC) or averaging properties calculated with different twisted boundary conditions (TBC) can significantly improve the extrapolation to the TL.

\section{Superfluid Neutron Matter: A Strongly Interacting Fermionic System}

\subsection{The variety of approaches in superfluid neutron matter}
\label{subsec:NM}
Neutron matter is a strongly interacting Fermi gas making for both intriguing physics as well as a system where known weak-interaction approximations break down. Historically, the strongly interacting nature of neutron matter gave rise to the study of the unitarity regime which was conceptualized as a model for the dilute neutron gas by Bertsch~\cite{Bertsch:1999} and Baker~\cite{Baker:1999}. This is the regime where a Fermi gas, while still dilute (the inter-particle distance is much larger than the range of the interaction) is very strongly interacting (the scattering length of the interaction diverges, $k_Fa\to\infty$)\cite{Giorgini:2008}. This ensures that all length scales associated with the interaction disappear making the two-body problem scale invariant and parameter-free. Today, unitarity is known to lie in the crossover separating the BCS regime of a Fermi gas from the Bose-Einstein Condensate (BEC) regime, aptly named the BCS-BEC crossover. For vanishingly small effective range $r_{\textrm{e}}$, the BCS-BEC crossover can be parametrized by the inverse of the coupling constant, $1/{k_Fa}$. Starting on the BCS side with $1/k_F a< 0$, one finds particles arranged in Cooper pairs bound with energy $\Delta$, also known as the pairing gap. Their binding increases with the coupling, reaching unitarity at $1/k_Fa=0$. Past unitarity lies the BEC regime, with $1/k_Fa > 0$, where the tightly bound Cooper pairs have become bosonic diatomic molecules and are condensed in the familiar BEC state. Neutron matter found in the inner crust of quiescent neutron stars, with its large $S$-wave scattering length $a_s\approx -18.5~\textrm{fm}$ and relatively small effective range $r_{\textrm{e}} \approx 2.7~\textrm{fm}$, is situated close to unitarity, on the BCS side -- a proximity that has led to connections between neutron matter and cold Fermi atoms \cite{Baker:1999, Strinati:2018, Carlson:2003, Heiselberg:2000, deMelo:1993, Gezerlis:2008, Regal:2003, Bourdel:2003} (for a generalization of the crossover for finite $r_\textrm{e}$ see \cite{Tajima:2019}). In the latter, the scattering length of the inter-atomic interaction and its sign can be changed by tuning an external magnetic field (Feshbach resonances \cite{Regal:2004, Bartenstein:2004, Chin:2010,Partridge:2005}) allowing one to navigate the BCS-BEC crossover and perform experiments at unitarity, where the details of the interaction are irrelevant and universal conclusions can be drawn. At the same time, the proximity of neutron matter to the unitary Fermi gas makes it the most strongly paired fermion superfluid system known in Nature with calculations and experiments suggesting pairing gaps peaking at 30 per cent of the Fermi energy~\cite{Gezerlis:2015}. The physics of these strongly-interacting superfluids is not understood nearly as well as its weakly-interacting counterparts.

The superfluid neutron matter of the inner crust does not exhibit the universal behavior expected from unitary gases entirely, despite its proximity to the unitarity regime. The reason behind this is its finite effective range \cite{Gezerlis:Novel}. The low-energy phenomena of an interacting gas depend only on the large scale characteristics of the interaction since higher energy is needed to probe finer details in the potential. This is the essence of the effective range expansion, where the physics of the 2-body phase shift, namely $\delta _0$ are captured by the scattering length $a$ and the effective range  $r_\textrm{e}$ in
\begin{equation}
    \cot{\delta_0} = -\frac{1}{a} + \frac{1}{2} r_\textrm{e}k^2 + \dots~. ~\label{eq:ef_range_exp}
\end{equation}
\begin{figure}[htp]
\begin{center}
\includegraphics[width=0.8\columnwidth,clip=]{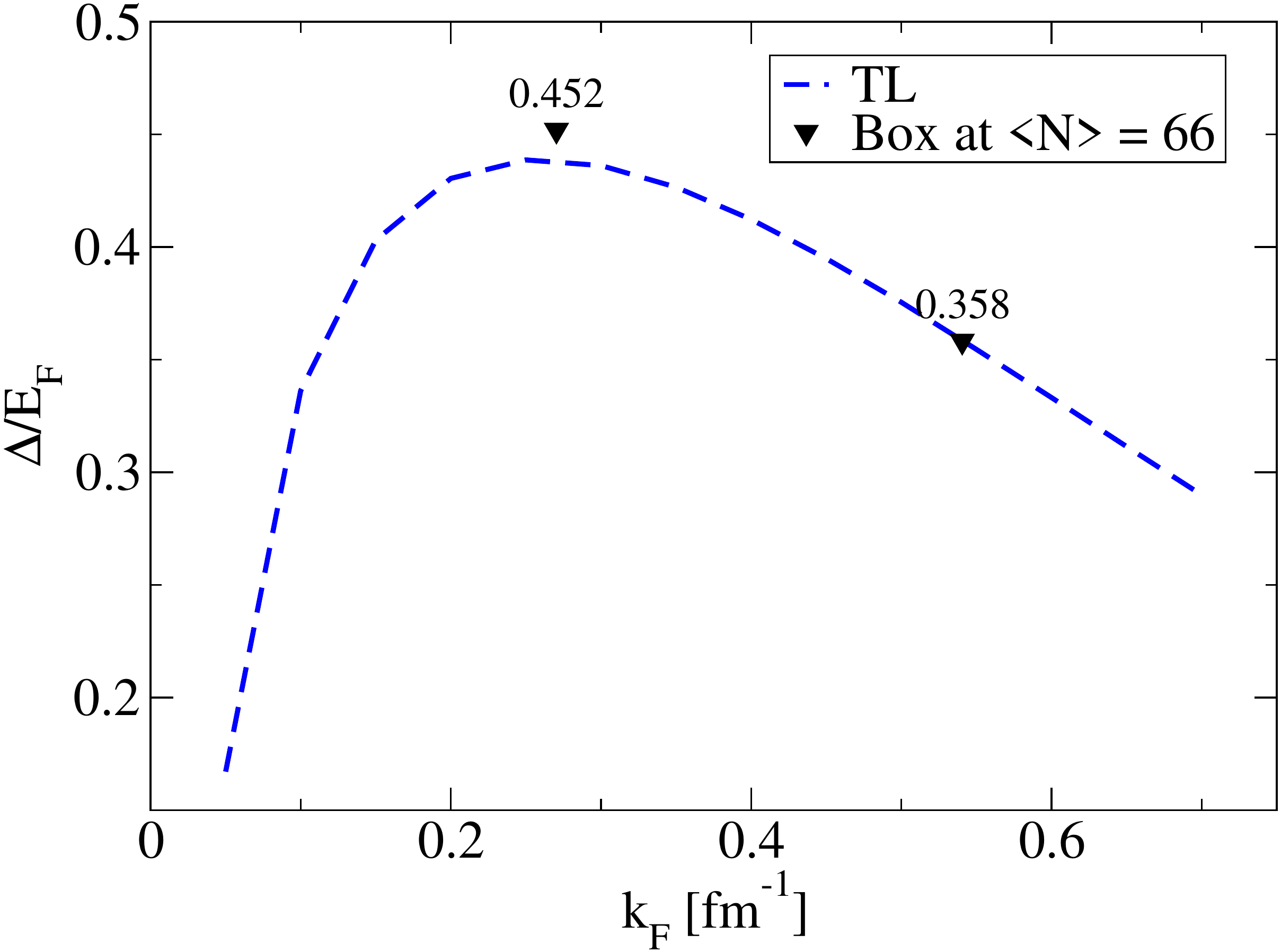}
\caption{The pairing gap at the TL divided by the Fermi energy as a function of the Fermi momentum $k_{\textrm{F}}$. Also plotted are the pairing gaps for $\left< N \right> =66$ at $k_\textrm{F}a=-10$ and $-5$.	\label{fig:DEf_kf}}
\end{center}
\end{figure}

The pairing responsible for superfluidity happens predominantly in momenta lying in a band centered at the Fermi level $k_\textrm{F}$ whose width increases with the pairing strength, as noted below. That allows us to investigate the range of validity of the effective range expansion by considering only momenta $k\approx k_\textrm{F}$. Using this, one can see that, in NM interacting through the $S$-channel, at $k_\textrm{F}a\approx-5$ the momentum dependent terms become of the order of $1/a$ already setting it apart from a unitary gas. As $k_\textrm{F}a$ increases further, the importance of the momentum-dependent terms becomes higher reaching $k_\textrm{F}a\approx -10$ where the presence of these terms becomes vital for the correct description of neutron matter. This is not far from the interpretation of higher-energy phenomena as probes for the finer-scale characteristics of the interaction. As the density increases, $k_\textrm{F}$ is pushed to higher momenta and so is the the band of momentum states relevant to pairing. With scattering in higher momentum states present, finer details of the potential become relevant distinguishing neutron matter from unitary gases. This can be seen in QMC calculations of the pairing gap in neutron matter and cold atoms \cite{Gezerlis:2008}. Another feature of the effective range is the initiation of an effective repulsion at higher densities. The positive second term in Eq.~(\ref{eq:ef_range_exp}) has a plus sign unlike the preceding negative term whose magnitude and sign are related to the ``attractiveness'' of the interaction. This reduction of the attractive interaction leaves its footprint at higher densities where the gap of the superfluid state (as a fraction of the Fermi energy, $E_F$) decreases after $k_\textrm{F}a \gtrsim$ -5, (see Fig.~\ref{fig:DEf_kf}) instead of saturating as expected from unitary gases. Finally the attractive $S$-channel of the NN interaction turns repulsive at $k_\textrm{F} \sim 1.5~\textrm{fm}^{-1}$~\cite{Stoks:1997} instigating the closure of the pairing gap in that channel at the corresponding densities~\cite{Hebeler:2007}. 

Having identified neutron matter's place in the BCS side of the BCS-BEC crossover, what follows will focus on that region. There, on the weak-coupling limit, $1/k_Fa \to -\infty$, the (mean-field) BCS theory of superfluidity yields a correct qualitative and quantitative description of pairing \cite{Gezerlis:2010} with an analytic expression for the pairing gap:
\begin{equation}
    \Delta _{\textrm{BCS}} = \frac{8}{e^2}\frac{\hbar ^2 k_F^2}{2m}\exp{\left(\frac{\pi}{k_F a}\right)}
\end{equation}
When including polarization effects, the above expression acquires an extra $k_Fa$-dependent factor which tends to $(1/4e)^{1/3}\approx 0.45$ at the limit of $1/k_Fa\to -\infty$~\cite{Schulze:2001, Gorkov:1961}. As one moves to strongly-paired superfluids, such approaches, while still valid qualitatively, fail to provide a quantitative description. The BCS description, while accurately predicting the two-body bound states that lie on the BEC side of the crossover, does not provide precise results in that regime. For these systems one needs to turn to beyond-mean-field approaches. These approaches are built by considering correlations neglected by mean-field theories. The inclusion of short- and long-range correlations by a summation of ladder diagrams in the context of self-consistent Green's function method results in a $\sim 0.75~\textrm{MeV}$ reduction of the pairing gap compared to the BCS value, and in a closure of the pairing gap at lower densities~\cite{Rios:2016}. Similar results were found by replacing the bare interaction with the $G$-matrix in the Bruckner theory to include screening effects~\cite{Cao:2006}. In-medium corrections to the effective interaction can also be included ``adiabatically'' by solving the Renormalization Group (RG) flow equations in an RG approach~\cite{Schwenk:2003} where polarization effects of the particle-hole kind result in a smaller gap compared to BCS. Similarly, reduced gaps are found when the particle-particle and particle-hole polarization effects are added by phenomenologically modifying the short-distance behavior of the bare fermion interaction~\cite{Wambach:1993}. A significant quenching of the gap is also found when treating short-range correlations by means of Correlated Basis Functions (CBF)~\cite{Mavromatti:2017}. The addition of three-body forces has also been explored in the context of chiral effective interactions where a reduced $^1S_0$ pairing gap and an increased $^3PF_2$ pairing gap were reported~\cite{Maurizio:2014}. Finally the complete quantum many-body problem can be solved using the Quantum Monte Carlo (QMC) family of stochastic approaches where the ground-state of a many-body Hamiltonian is identified given a suitable trial wave-function~\cite{Gandolfi:2008, Gezerlis:2008}. Calculations of the $^1S_0$ pairing gap by the means of QMC simulations, which first resulted in such a reduction of the gap as compared to the BCS value. An important distinction of the QMC methods is that one is obliged to work with finite systems and, therefore, calculations of quantities of infinite matter require the extra step of extrapolation, i.e., dealing with the FSE (see section~\ref{sec:FSE}). These techniques use the Rayleigh-Ritz principle to find a state with minimum energy combined with an imaginary-time propagation to purify that state of any excited-state contributions. When fermionic, the imaginary-time evolution of such states contains the additional complication of the fermion-sign problem arising from anti-symmetric states losing their anti-symmetry due to the statistical sampling inherent in the stochastic method (for a review of QMC methods see Ref~\cite{Foulkes:Book}). With the advancements of the QMC approach over the few past decades having rendered the fermion-sign problem under control, QMC results can be treated as exact yielding an error margin of typically $~\sim 1\%$~\cite{Gezerlis:Novel}. For a more detailed review on various approaches for the calculation of the pairing gap in neutron matter see Ref.~\cite{Gezerlis:Novel}.

\subsection{The BCS and PBCS Theories for Neutron Matter}
\label{sec:BCS}
The BCS theory introduced in 1957 by Bardeen, Cooper, and Schrieffer, describes superfluid (or superconducting) states as a result of an instability of the normal state in fermionic gases, with an attractive interaction, at low temperatures. In BCS, this instability, present below a transition temperature, instigates a formation of Cooper pairs. The attractive interaction responsible for the instability of the normal state and the eventual formation of the condensate of pairs, as the temperature is further decreased, can come from various sources. In the original formulation of the theory, in the context of superconductivity in metals, the origin of the attractive interaction was the presence of an ionic lattice which allows for an effective interaction between electrons via the exchange of a phonon. In NM the nature of the attractive interaction is different, as it originates from the $S$-wave component of the NN interaction which is attractive at densities encountered in the inner crust of cold NSs. As mentioned in subsection~\ref{subsec:NM}, the strong nature of this interaction which is portrayed by its large scattering length $a_\textrm{s}\approx -18.5~\textrm{fm}$, makes NM one of the most strongly correlated superfluids encountered in Nature. 

The BCS theory describes superfluidity on a mean-field level and so it is not expected to give accurate quantitative results for strongly interacting superfluids such as NM. However, one can use the BCS theory and its straightforward treatment of pairing to produce qualitative results and offer guidance to more accurate methods. Specifically, one can use results from a BCS treatment to help \textit{ab initio} techniques, applied to finite systems, extrapolate to the thermodynamic limit (TL). For more details on the assistance that BCS theory can provide to these extrapolations see section~\ref{sec:FSE}. Here we present a brief introduction on the BCS theory, and its particle-conserving variation, namely PBCS theory, applied to NM. 

One can use the BCS theory to describe a finite part of the bulk of a pure NM superfluid by enclosing a system of $N_0$ neutrons in a finite volume, e.g. a cubic box of length $L$, under PBC. The PBC require that the modulus of the wavefunction be the same on opposite sides of the box allowing for a complex phase difference between the values of the wave-function on  the same points:
\begin{align}
    \left|\psi \left(\mathbf{r}_1+L\hat{x}, \mathbf{r}_2, \dots , \mathbf{r}_{N_0}\right)\right|^2 = \left|\psi \left(\mathbf{r}_1, \mathbf{r}_2, \dots , \mathbf{r}_{N_0}\right)\right|^2 \\
    \psi \left(\mathbf{r}_1+L\hat{x}, \mathbf{r}_2, \dots , \mathbf{r}_{N_0}\right) = e^{i\theta _x}\psi \left(\mathbf{r}_1, \mathbf{r}_2, \dots , \mathbf{r}_{N_0}\right)
\end{align}
Following the literature, we name this phase difference ``twist'' to avoid the polysemous word ``phase''. The choice of PBC is mandated by the translational symmetry that characterizes all quantities of infinite matter, and it gives rise to the familiar single-particle spectrum
\begin{align}
    \epsilon _\mathbf{k} &= \frac{\hbar ^2}{2m} \left|\mathbf{k}\right| ^2~, ~\label{eq:sp} \\
    \mathbf{k} &= \frac{2\pi}{L}\left(\mathbf{n}+\frac{\boldsymbol{\theta}}{2\pi}\right),\quad n_i=0,\,\pm 1,\, \pm 2,\,\dots ~. \label{eq:BC}
\end{align}
The twist angle $\boldsymbol{\theta}$ is not an observable of the infinite system: at the limit of $L\to \infty$ it drops out of the problem. Therefore, it is typical to assume untwisted PBC,i.e., $\boldsymbol{\theta}=0$. However, ways to manipulate this extra degree of freedom, and bring finite systems closer to TL, have been developed. In section~\ref{sec:FSE} we will use special twist-angles as well as a twist-averaging approach, where we average the quantities of a finite system over the twist angle, to make the finite system a better approximation of infinite matter. For the rest of this section we restrict our BC to the untwisted $\boldsymbol{\theta} = 0$ case,i.e., PBC.

When studying the effects of pairing in NM, it is customary to isolate only the interaction responsible for the pairing and ignore normal state interactions, hence describing the system using the so-called pairing Hamiltonian:
\begin{equation}
\hat{H} = \sum_{\mathbf{k} \sigma} \epsilon _\mathbf{k} \hat{c}_{\mathbf{k} \sigma}^\dagger \hat{c}_{\mathbf{k} \sigma} + \sum_{\mathbf{k}\mathbf{l}} \left<\mathbf{k}\right|V\left|\mathbf{l}\right> \hat{c}_{\mathbf{k} \uparrow}^\dagger\hat{c}_{-\mathbf{k} \downarrow}^\dagger\hat{c}_{-\mathbf{l} \downarrow}\hat{c}_{\mathbf{l} \uparrow} ~,\label{eq:Hamiltonian}
\end{equation}
where $\hat{c}_{\mathbf{k}\sigma}^{\dagger},\, \hat{c}_{\mathbf{k}\sigma}$ are fermionic creation and annihilation operators, respectively, creating or annihilating free single-particle states with momentum $\mathbf{k}$ and spin $\sigma$ in a cubic box under PBC. Many state-of-the-art potentials have been used to model the NN interaction such as the CD-Bonn potential~\cite{Machleidt:1996}, the Nijmegen I and II potentials~\cite{Stoks:1994}, the Argonne family of potentials (AV4, AV8, etc)~\cite{Gezerlis:2010,Gandolfi:2008}, chiral potentials (NLO, N$^2$LO, and N$^3$LO)\cite{Gezerlis:2014}, etc. At low energies, where the details of the interaction do not matter, any potential that reproduces the $^1S_0$ scattering length and effective range of NM should produce identical results, as discussed already in section~\ref{subsec:NM}. In this light, we choose the simplest path and we model the NN interaction using the purely attractive P{\"o}schl-Teller~(PT) potential,
\begin{equation}
    V(r) = - \frac{\hbar}{m_n}\frac{\lambda (\lambda - 1)\beta ^2}{\cosh^2{(\beta r})} ~ ,	\label{eq:PT}
\end{equation}
\begin{figure}[htp]
\begin{center}
\includegraphics[width=0.8\columnwidth,clip=]{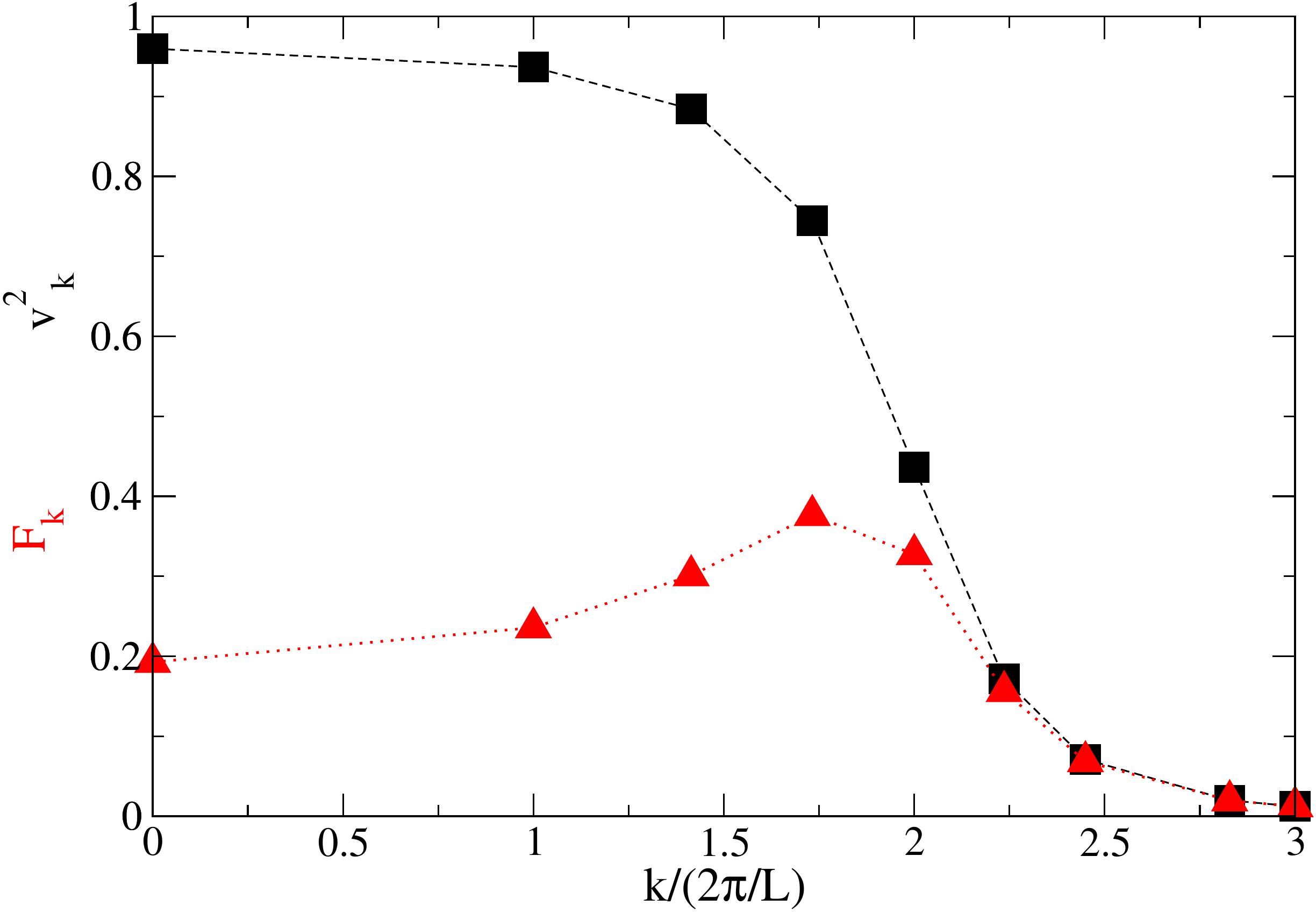}
\caption{The occupation probability distribution and the condensation amplitude for $\left<N\right>=66$ and $k_\textrm{F}a=-10$.	\label{fig:v2k_Fk}}
\end{center}
\end{figure}
where the parameters $\lambda$ and $\beta$ are adjusted to reproduce the scattering length and effective range of the $S$-wave of the NN interaction. The matrix element in Eq.~(\ref{eq:Hamiltonian}) is that of the $^1S_0$ channel of this potential,
\begin{align}
V_0 (k,k') &= \int _0 ^\infty dr r^2 j_0(kr)V(r)j_0(k'r) ~.
\end{align}
It is worth noting that assuming a purely attractive potential, such as the PT potential, we ignore the repulsive core of the NN interaction but, again consistent with the shape independence outlined in section~\ref{subsec:NM}, for low-density NM, the exact form of the potential is irrelevant and the results produced by all the potentials mentioned above should be identical~\cite{Gezerlis:2015, Gezerlis:2010}.

\subsubsection{Even particle-number superfluid}
The BCS theory describes the ground-state of a superfluid with an even number of particles $N_0$ as a coherent state of $N_0/2$ pairs of time-reversed states:
\begin{equation}
\left| \psi _{\textrm{BCS}} \right> = \prod_{\mathbf{k}} \left(u_{\mathbf{k}} + v_{\mathbf{k}} \hat{c}_{\mathbf{k}\uparrow}^{\dagger} \hat{c}_{-\mathbf{k}\downarrow}^{\dagger}\right)\left|0\right> ~.\label{eq:groundstateBCS}
\end{equation}
The state $\ket{0}$ stands for the vacuum while the distributions $v_{\mathbf{k}}$ and $u_{\mathbf{k}}$ represent the probability amplitude of finding or not finding, respectively, a pair of states $\mathbf{k}\uparrow,\,-\mathbf{k}\downarrow$, and so they are subject to the normalization
\begin{equation}
    v_{\mathbf{k}}^2 + u_{\mathbf{k}}^2 = 1~. \label{eq:vu_norm}
\end{equation}
Furthermore, with $v_{\mathbf{k}}^2$ representing the probability of finding a pair characterized by momentum $\mathbf{k}$, it is connected to the average number of particles,
\begin{equation}
    \left<\hat{N}\right>=\sum_{\mathbf{k}} 2v_{\mathbf{k}}^2=N_0~. \label{eq:v2k}
\end{equation}
\begin{figure}[htp]
\begin{center}
\includegraphics[width=0.8\columnwidth,clip=]{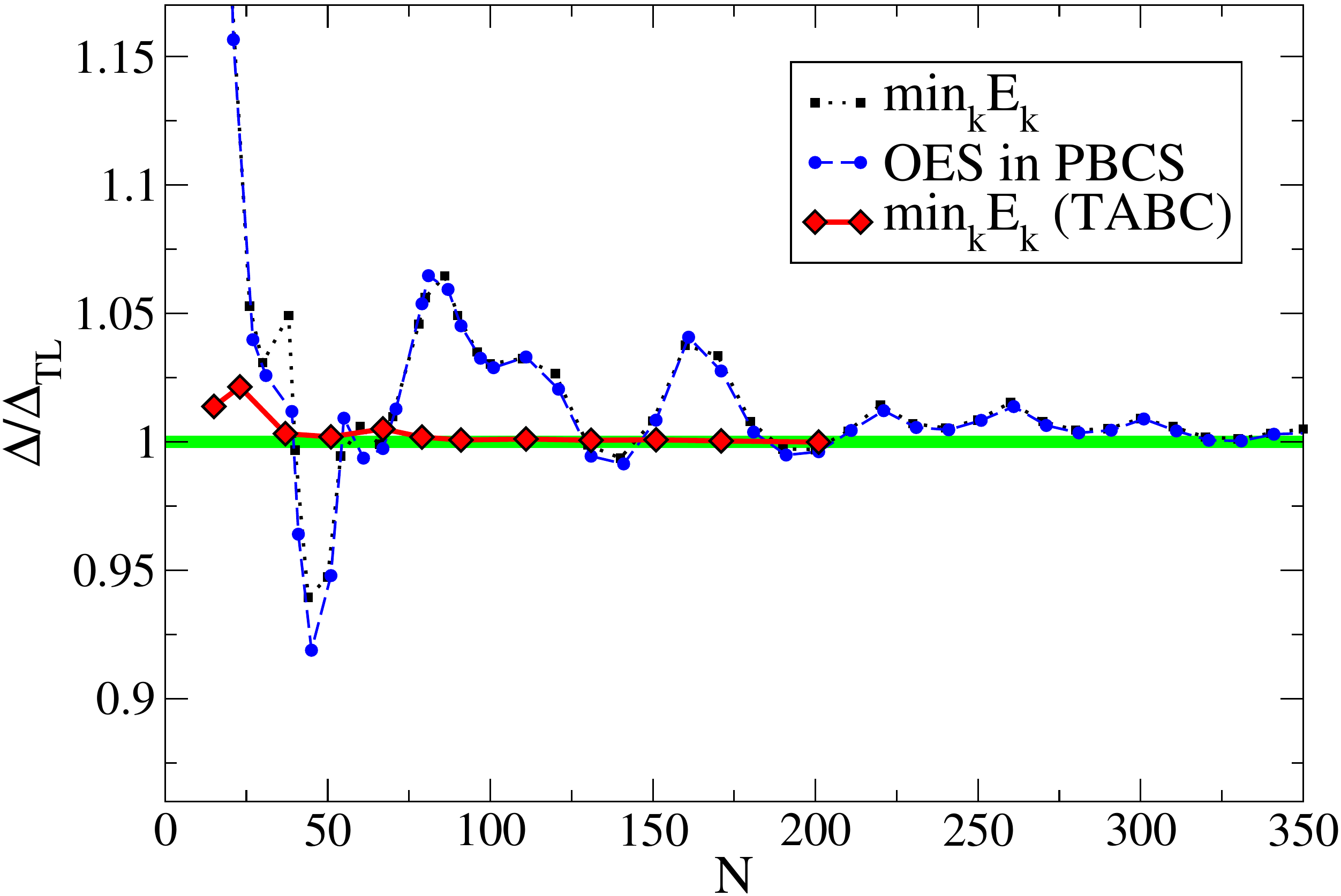}
\caption{The minimum of the quasi particle excitation energy and the OES in PBCS under PBC for $k_\textrm{F}a=-10$. The minimum of quasiparticle excitation energy under TABC is shown in red. \label{fig:D}}
\end{center}
\end{figure}
This average is to be understood in the context of a grand canonical ensemble: we enforce the particle-number conservation in the box only on average. In fact, the BCS ground state in Eq.~(\ref{eq:groundstateBCS}) is defined in a grand canonical ensemble and so it describes a superlfuid with an \textit{average} particle-number equal to $N_0$. Therefore, the probability distributions $v_\mathbf{k}^2$ and $u_\mathbf{k}^2$, which characterize the pair occupation of the condensate in $\mathbf{k}$-space, should be such that they minimize the free energy of the BCS state,
\begin{equation}
    F_\textrm{BCS}=\bra{\psi_\textrm{BCS}}\hat{H}-\mu\hat{N}\ket{\psi _\textrm{BCS}}=\sum_\mathbf{k}2v_\mathbf{k}^2\xi_\mathbf{k} +\sum_{\mathbf{k}\mathbf{l}}V_{\mathbf{k}\mathbf{l}}v_\mathbf{k}u_\mathbf{k}v_\mathbf{l}u_\mathbf{l}~,
\end{equation}
where $\xi_\mathbf{k}=\left(\epsilon_k-\mu\right)$ and $\mu$ is the chemical potential. This minimization yields the gap equation:
\begin{equation}
\Delta (k) = -\frac{2\pi}{L^3} \sum_{k'} M(k') V_0(k,k')\frac{\Delta(k')}{E(k')} 	~, \label{eq:Gap1BCSSwave}
\end{equation} 
where every momentum-dependent quantity is replaced by an angle-averaged version. This is because only the first term of the corresponding Laplace series of each quantity couples with the $S$-wave of the potential which reflects the fact that momentum states scattered by an $S$-wave interaction can only change in magnitude and not in direction. Thus, momentum shells can be defined, each containing momenta corresponding to the same $\mathbf{k}$-magnitude which are treated as identical by the interaction. The population of the shell corresponding to $|\mathbf{k}|=k$ is denoted by $M(k)$ and is used in the discrete sums over $\mathbf{k}$-space to allow for a one-dimensional sum over the discrete $\mathbf{k}$-magnitudes instead of the entire three-dimensional momentum space. The populations of the different momentum shells depends on the twist of the BC as well, which we have taken to be the trivial $\boldsymbol{\theta}=0$ in this section. See section~\ref{sec:FSE} for the dependence of $M(k)$ on $\boldsymbol{\theta}$ and how it affects the FSE. Here the gap function $\Delta_\mathbf{k}$ is the binding energy of a pair with momentum $\mathbf{k}$ and the energy needed to break a pair and create an excitation is the quasi-particle excitation energy $E_\mathbf{k}$. It is defined as 
\begin{equation}
    E_\mathbf{k}=\sqrt{\xi^2_\mathbf{k}+\Delta^2_\mathbf{k}}~, \label{eq:exc}
\end{equation}
\begin{figure}[htp]
\begin{center}
\includegraphics[width=0.8\columnwidth,clip=]{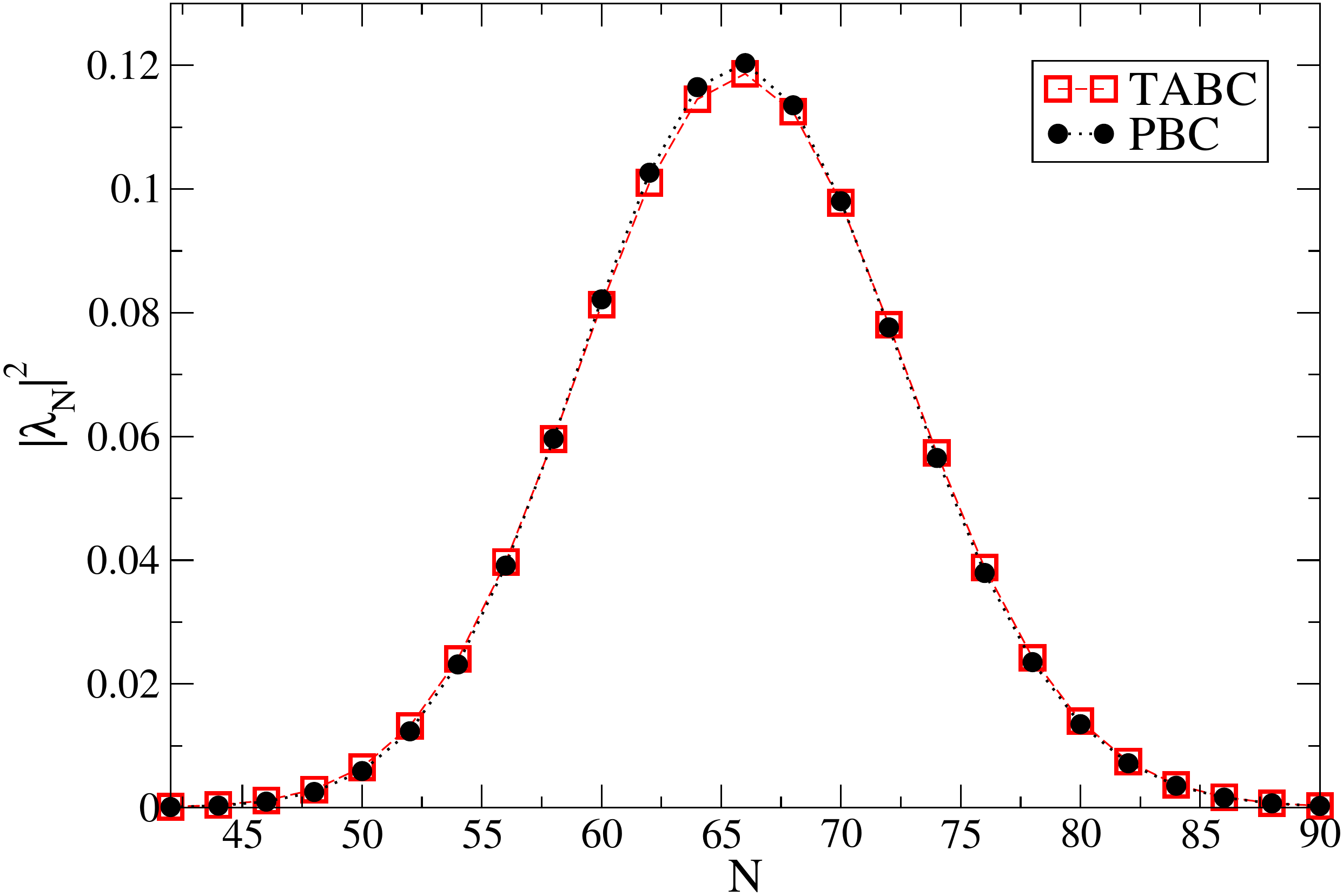}
\caption{The probability of finding a PBCS ground state with $N$ particles projected out of a BCS ground state with $\left<N\right>=66$ particles.}	\label{fig:R00}
\end{center}
\end{figure}
and it defines the occupation probability distributions, 
\begin{align}
v_{\mathbf{k}}^2 &= \frac{1}{2}\left(1-\frac{\xi _{\mathbf{k}}}{E _{\mathbf{k}}}\right)  ~, \label{eq:v2kdefinition}\\ 
u_{\mathbf{k}}^2 &= \frac{1}{2}\left(1+\frac{\xi _{\mathbf{k}}}{E _{\mathbf{k}}}\right)  ~. \label{eq:u2kdefinition}
\end{align}
Given these, Eq.~(\ref{eq:v2k}) becomes
\begin{equation}
    \left<\hat{N}\right> = \sum _{k} M(k) \left(1-\frac{\xi(k)}{E(k)}\right)	 ~, 	\label{eq:Gap2BCSSwave}
\end{equation}
which again involves angle-averaged quantities and the population of the $\mathbf{k}$-shells. Equations~(\ref{eq:Gap1BCSSwave})~\&~(\ref{eq:Gap2BCSSwave}) are called the \textit{BCS gap equations} and their solution defines the distributions $v_{\mathbf{k}}$ and $u_\mathbf{k}$ through Eqs.~(\ref{eq:v2kdefinition})~\&~(\ref{eq:u2kdefinition}) which in turn will define the BCS energy of the superfluid in terms of the angle-averaged distributions:
\begin{equation}
    E^{\textrm{BCS}}_{\textrm{even}}  (N) = \sum_{k} M(k) \epsilon _{k} 2v^2(k) + \frac{4\pi}{L^3} \sum_{k k'} M(k)M(k') V_0(k,k') u(k)v(k) u(k') v(k') ~.	\label{eq:SwaveEnergyBCS}
\end{equation}

It is worth noting that the normal state can be written in the form of the BCS ground state with $v_\mathbf{k}=1$ for $|\mathbf{k}|\leq k_\textrm{F}$ and $v_\mathbf{k}=0$ otherwise (this corresponds to the $\Delta_\mathbf{k}=0$ solution). This means that when minimizing the free energy of the BCS state, i.e., Eq.~(\ref{eq:groundstateBCS}), with respect to the distribution $v_\mathbf{k}$, the normal state is a viable candidate. In other words, if the solution of the BCS gap equations yields solutions other than the normal state solution, then a pair condensate yields lower free energy than the normal state. This is an equivalent statement of the pairing instability proposed in the initial BCS theory. 

As seen in Fig.~\ref{fig:v2k_Fk}, the distributions $v_\mathbf{k}^2$ is smeared over $\mathbf{k}$-space, compared to the Fermi distribution. This is a consequence of pairing: taking $\Delta_\mathbf{k}\to 0$ in Eq.~(\ref{eq:v2kdefinition}) one can retrieve the Fermi distribution, i.e., a free Fermi gas. This is demonstrated by the condensation amplitude
\begin{equation}
    F_\mathbf{k} = v_\mathbf{k}u_\mathbf{k}~, \label{eq:condensation_amplitude}
\end{equation}
which can also be seen in Fig.~\ref{fig:v2k_Fk}. The product $v_\mathbf{k}u_\mathbf{k}$ is non-zero only when $v_\mathbf{k}$ and $u_\mathbf{k}$ are simultaneously non-zero. Hence, the spread of the condensation amplitude is strongly related to pairing with a wide spread $F_\mathbf{k}$ characterizing a strongly paired superfluid and a normal state yielding an infinitely sharp $F_\mathbf{k}$ around $|\mathbf{k}|=k_\textrm{F}$.

A key property of superfluidity is the energy gap in the quasi-particle excitation spectrum, e.g., see Fig.~\ref{fig:v2k_Fk}. The minimum energy required to break a pair and create an excitation is called the pairing gap and it is defined as
\begin{equation}
    \Delta _\textrm{BCS} = \textrm{min}_\mathbf{k} E_\mathbf{k}~. \label{eq:pairinggap}
\end{equation}
Figure~\ref{fig:D} shows the pairing gap as a function of $\left<N\right>$. Apart from the familiar oscillations seen in all intensive quantities of finite systems (see section~\ref{sec:FSE}), the pairing gap suffers from additional FSE due to its definition as a minimum of the discrete spectrum $E_\mathbf{k}$. In more detail, the quantization of momenta, imposed on finite systems by their BC, ensures that all functions of momentum are a discrete version of their TL-counterparts. Quantities defined as the minima of such discrete functions, in the manner of Eq.~(\ref{eq:pairinggap}), will have to compromise with the lowest available point. The position of this available minimum might change to a neighboring $\mathbf{k}$-state as $\left<N\right>$ changes and thus make the $\left<N\right>$-dependence appear more random.

The definition of the pairing gap, namely Eq.~(\ref{eq:pairinggap}), makes it hard to compare the BCS approach to superfluidity with other techniques. A prime example of this difficulty is QMC simulations. Drawing from the odd-even mass staggering in nuclear physics, one can define a quantity similar to the pairing gap which, at the TL, is identical to the pairing gap:
\begin{equation}
    \Delta = E(N) - \frac{1}{2}\left(E(N+1)+E(N-1)\right)~, \label{eq:OES}
\end{equation}
where $N$ is an odd number of particles since it provides better decoupling of the result from the underlying mean-field theory~\cite{Duguet:2:2001}, i.e., in this case, BCS. The odd-even staggering (OES) in neutron superfluidity has been demonstrated to reproduce the BCS pairing gap for systems far from the TL as well~\cite{Palkanoglou:2020}. The OES is a quantity which is readily available to a variety of \textit{ab initio} approaches to superfluidity, it can be used to compare them and quantify differences between them. To have access to OES in the BCS theory via Eq.~(\ref{eq:OES}) one needs an expression for the energy of a system with $N$ particles. As discussed above, the BCS treatment breaks the particle-conservation of the pairing Hamiltonian resulting in a state defined in a grand canonical ensemble with a constant \textit{average} particle-number. Hence, the BCS state can be understood as a superposition of eigenstates of the number operator, i.e., states describing superfluids with fixed particle number:
\begin{equation}
    \ket{\psi _{\textrm{BCS}}} = \sum_N \lambda _N \ket{\psi _N}~. \label{eq:N_exp}
\end{equation}

The spectrum of eigenstates $|\lambda_N|^2$ is sharply peaked around $N=N_0$, as seen in Fig.~\ref{fig:R00}, with a relative spread proportional to $N_0^{-1/2}$. In other words, the main contribution in Eq.~(\ref{eq:N_exp}) largely comes from the state $\ket{\psi _{N_0}}$ and that leads to the idea of the projected BCS (PBCS) theory where the ground-state of a superfluid with an exact (even) particle-number $N_0$ is described by $\ket{\psi _{N_0}}$. The less sharp the peak of $|\lambda_N|^2$  the more particle-conserving the BCS ground state $\ket{\psi _\textrm{BCS}}$ is and the closer PBCS is to BCS. It follows that for strongly interacting superfluids, of which NM is a perfect specimen, the particle-conserving PBCS theory is expected to agree with the non-particle-conserving BCS theory.

In PBCS theory one projects out of the sum in Eq.~(\ref{eq:N_exp}) the state corresponding to $N_0=\left<N\right>$, namely $\left|\psi_{N_0}\right>$. To execute this projection we need to allow $v_\mathbf{k}$ and $u_\mathbf{k}$ to take complex values and with the BCS ground state being a coherent state of pairs, the complex phase of $v_\mathbf{k}$ and $u_\mathbf{k}$ does not depend on $\mathbf{k}$. Furthermore, having the freedom to define the BCS ground state up to an overall complex phase leads to 
\begin{equation}
    \left| {\psi _{\textrm{BCS}}}_\phi \right> = \prod_{\mathbf{k}} \left(u_{\mathbf{k}} + v_{\mathbf{k}}e^{i\phi} \hat{c}_{\mathbf{k}\uparrow}^{\dagger} \hat{c}_{-\mathbf{k}\downarrow}^{\dagger}\right)\left|0\right> ~.\label{eq:groundstateBCS_phi}    
\end{equation}
The infinite product in Eq.~(\ref{eq:groundstateBCS_phi}) generates an infinite sum in which each term comes with as many factors of $e^{i\phi}$ as pair creation operators $\hat{c}_{\mathbf{k}\uparrow}^{\dagger} \hat{c}_{-\mathbf{k}\downarrow}^{\dagger}$. Therefore, the term creating $N_0/2$ pairs can be projected out by integrating the rest of the terms over a integer number of periods in $\phi$:
\begin{equation}
    \left|\psi _N\right> = \frac{1}{R_0^0}\int_{0}^{2\pi} \frac{d\phi}{2\pi} e^{-i\frac{N}{2}\phi} \prod_{\mathbf{k}} \left(u_{\mathbf{k}} + e^{i\phi} v_{\mathbf{k}} \hat{c}_{\mathbf{k}\uparrow}^{\dagger} \hat{c}_{-\mathbf{k}\downarrow}^{\dagger}\right) \left|0\right>	~ ,	\label{eq:PBCSstate}
\end{equation}
where the normalization $R_0^0$ is one of the residuum integrals:
\begin{equation}
 R_n^m(\mathbf{k}_1 \mathbf{k}_2 \dots \mathbf{k}_m;N) = \int _0^{2\pi} \frac{d\phi}{2\pi} e^{-i(\frac{N}{2}-n)\phi} \prod _{\mathbf{k} \neq \mathbf{k}_1, \mathbf{k} _2, \dots \mathbf{k} _m} \left( u_{\mathbf{k}}^2 + e^{i\phi} v_{\mathbf{k}}^2\right)	~.	
	\label{eq:ResInt}
\end{equation}
The residuum integrals are related to the probabilities of arrangements of pairs in $\mathbf{k}$-space. For example, $R_0^0$ is the sum of the probabilities of all possible pair-arrangements in $\mathbf{k}$-space and as such it is the probability of the realization of the state in Eq.~(\ref{eq:PBCSstate}) and, therefore, its normalization. This makes $R_0^0$ equal to the $\left|\lambda_N\right|^2$ presented in Fig.~\ref{fig:R00}. A detailed derivation of the PBCS theory can be found in Ref.~\cite{Palkanoglou:2020}. The energy corresponding to the state in Eq.~(\ref{eq:PBCSstate}) is
\begin{equation}
E^{\textrm{PBCS}}_{\textrm{even}}(N) = \sum_{k} M(k) \epsilon _{k} 2v^2(k) \frac{R_1^1(k)}{R_0^0}+ 	\frac{4\pi}{L^3} \sum_{k k'}  M(k)M(k') V_0(k,k') u(k)v(k) u(k')v(k') \frac{R_1^2(kl)}{R_0^0}	~ ,	 \label{eq:SwaveEnergyPBCS}
\end{equation}
where we have again used the angle-averaged distributions. Owing to the projection, $E^{\textrm{BCS}}_{\textrm{even}}$ describes systems with a fixed particle number and, therefore, it can be used to calculate the gap using the OES. The $v_\mathbf{k}$ and $u_\mathbf{k}$ distributions in Eq.~(\ref{eq:SwaveEnergyPBCS}) are the ones in Eqs.~(\ref{eq:v2kdefinition})~\&~(\ref{eq:u2kdefinition}), respectively, since they originate from the already optimized, BCS ground-state in Eq.~(\ref{eq:groundstateBCS_phi}). One could re-optimize the PBCS state and find the distributions $v_\mathbf{k}$ and $u_\mathbf{k}$ that minimize the energy in Eq.~(\ref{eq:SwaveEnergyPBCS}). This corresponds to the FBCS approach~\cite{Dietrich:1964} of which BCS is a saddle point approximation valid for $2 \sqrt{\sum F_\mathbf{k}^2} \gg 1$. Following the preceding discussion on the condensation amplitude $F_\mathbf{k}$, it is clear that FBCS for NM should be qualitatively and quantitatively very close to BCS and, \textit{a fortiori}, to PBCS, allowing one to use the latter when probing quantities such as the OES.

\subsubsection{Odd particle-number systems}

The BCS ground state describes a condensation of pairs and as such it is not suitable for systems with an odd number of particles. With the energy of an odd particle-numbered system being central to the OES (cf. Eq.~(\ref{eq:OES})), a variation of the BCS ground state is essential for the calculation of the gap. This is the blocked BCS state,
\begin{equation}
\left|\psi _{\textrm{BCS}}^{\mathbf{b} \gamma}\right> = \hat{c}_{\mathbf{b}\gamma}^\dagger \prod_{\mathbf{k} \neq \mathbf{b}} \left(u_\mathbf{k} + v_\mathbf{k} \hat{c}_{\mathbf{k} \uparrow}^\dagger \hat{c}_{-\mathbf{k} \downarrow}^\dagger\right)\left| 0\right> ~,	\label{eq:bBCS} 
\end{equation}
where the extra, unpaired, particle occupies a momentum state $\mathbf{b}$ blocking it from the rest of the underlying condensate. The optimization of this state gives rise to its own blocked BCS equations
\begin{align}
    \Delta _\mathbf{k} &= -\frac{1}{2} \sum_{\mathbf{k}'\neq \mathbf{b}} V_0(\mathbf{k},\mathbf{k}')\frac{\Delta _\mathbf{k}'}{E_\mathbf{k}'} 	 ~,\label{eq:bGap1BCS} \\
\left<\hat{N}\right>  &= \sum _{\mathbf{k}\neq \mathbf{b}} \left(1-\frac{\xi _\mathbf{k}}{E_\mathbf{k}}\right)	~ .	\label{eq:bGap2BCS}
\end{align}
where the quantities $\Delta (\mathbf{k})$, and $E (\mathbf{k})$ retain their definitions as the gap distribution and quasi-particle excitation energy, respectively. The minimum of $E(\mathbf{k})$ is defined as the BCS pairing gap for the odd particle-number systems. Following the same principles as the ones used for the projection in the even particle-number systems, one can define a fixed particle-number state for odd $N$. This is done by projecting out of the underlying condensate an $N-1$ particle-conserving state,
\begin{equation}
    \left| \psi _{N}^{\mathbf{b} \gamma} \right> =\frac{1}{R_0^1(\mathbf{b})}\hat{c}_{\mathbf{b}\gamma}^\dagger  \int_{0}^{2\pi} \frac{d\phi}{2\pi} e^{-i\frac{N-1}{2}\phi}  \prod_{\mathbf{k} \neq \mathbf{b}} \left(u_{\mathbf{k}} + e^{i\phi} v_{\mathbf{k}} \hat{p}_{\mathbf{k}}^{\dagger}\right) \left|0\right>	~.	\label{eq:bPBCSstate}
\end{equation}
Finally, the energy of an odd number of particles in PBCS is the energy that corresponds to the state in Eq.~(\ref{eq:bPBCSstate}):
\begin{equation}
    E^{\textrm{PBCS}}_{\textrm{odd}}(b;N) = \sum_{k\neq b} M(k) \epsilon _{k} 2v^2(k) \frac{R_1^2(bk)}{R_0^1(b)} +  \epsilon _b + \frac{4\pi}{L^3} \sum_{k, k'\neq b} M(k)M(k')  V_0(k,k') u(k)v(k) u(k')v(k') \frac{R_1^3(bkl)}{R_0^1(b)}	~, \label{eq:bSwaveEnergyPBCS}
\end{equation}
where, similarly to Eq.~(\ref{eq:SwaveEnergyPBCS}),  we use the angle-averaged distributions along with the a slight abuse of notation: the blocking of the momentum state $\mathbf{b}$ does not block an entire shell but rather reduces its occupancy by $1$. The unpaired particle's contribution in $E^{\textrm{PBCS}}_{\textrm{odd}}$ is that of a free particle since all normal state interactions are excluded in the pairing Hamiltonian (cf. Eq.~(\ref{eq:Hamiltonian})). However, the presence of the extra particle is ``felt'' by the condensate via the blocking of the momentum state occupied by the particle. The energy in Eq.~(\ref{eq:bSwaveEnergyPBCS}) depends on the blocked state $\mathbf{b}$ and so when used in an OES prescription, the right $\mathbf{b}$ should be chosen to minimize $E_{odd}^{PBCS}$. This is because the odd particle-number system in OES is to be understood as the lowest quasi-particle excitation of its even particle-number vacuum. For the sake of completeness, the corresponding energy of an odd particle-numbered superfluid in BCS is:
\begin{equation}
    E^{\textrm{BCS}}_{\textrm{odd}}(b;N) = \sum_{k\neq b} M(k) \epsilon _{k} 2v^2(k)  +  \epsilon _b + \frac{4\pi}{L^3} \sum_{k k'\neq b} M(k)M(k')  V_0(k,k') u(k)v(k) u(k')v(k') 	~. \label{eq:bSwaveEnergyBCS}
\end{equation}

\subsubsection{The Thermodynamic Limit}

The treatment of the TL in BCS is straightforward. A system at the TL corresponds to the limiting case $L \to \infty$ of the periodic box used for finite systems. In that limit, the discretization of the momenta becomes infinitely fine, turning all previously discretized distributions into continuous functions of $\mathbf{k}$, and the sums in the corresponding equations into integrals. Hence the gap equations describing the system at the TL are:
\begin{align}
\Delta(k) &= -\frac{1}{\pi} \int_{0}^\infty dk' (k')^2 V_0(k,k') \frac{\Delta (k')}{E(k')} ~, \label{eq:TL_SwaveGap1BCS} \\
\frac{\left<N\right>}{L^3} = n &= \frac{1}{2\pi ^2} \int_{0}^\infty dk k^2 \left(1-\frac{\xi (k)}{E(k)}\right) ~ . \label{eq:TL_SwaveGap2BCS} 
\end{align}
where $n$ is the number density, and $\Delta (k)$ and $E (k)$ are the continuous version of the gap distribution and quasi-particle excitation energy, respectively. Hence, a BCS pairing gap can be again defined in an identical way to the even and odd particle-number systems:
\begin{equation}
    \Delta_\textrm{BCS}=\textrm{min}_\mathbf{k} E(\mathbf{k})~. \label{eq:gapTL}
\end{equation}

From the discussion on the spread around the peak of the eigenstate spectrum $\lambda_N^2$ it follows that for $N\to \infty$, i.e., at the TL, the peak is infinitely sharp and the particle number projection loses its meaning and so does the distinction between even and odd particle numbers.

\begin{figure}[htp]
    \begin{subfigure}{.48\textwidth}
      \centering
        \includegraphics[width=\columnwidth,clip=]{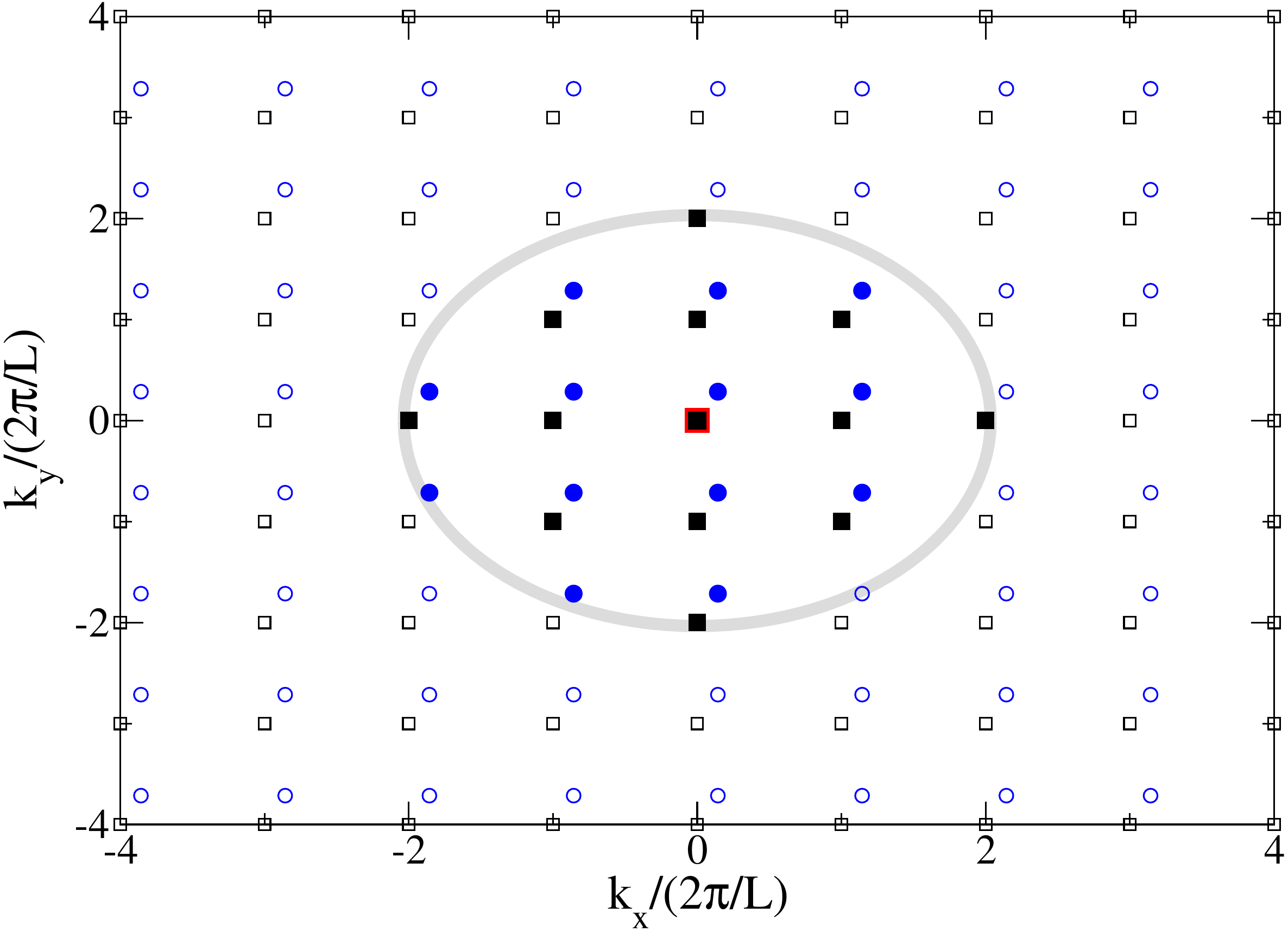}
        \caption{\label{fig:kspace}}
    \end{subfigure}
    \hfill
    \begin{subfigure}{.48\textwidth}
      \centering
        \includegraphics[width=\columnwidth,clip=]{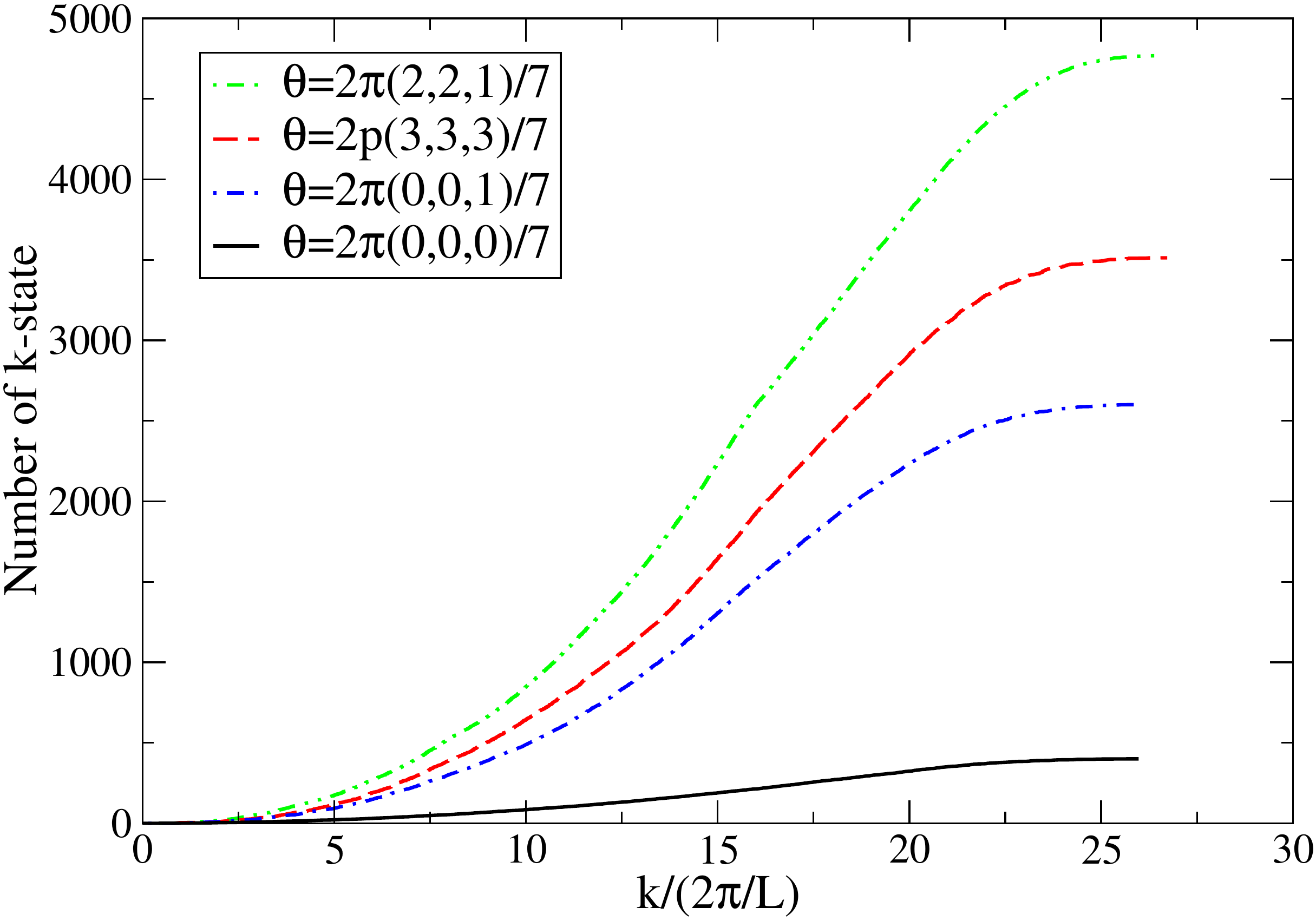}
       \caption{\label{fig:m_twist}}
    \end{subfigure}
    \vspace{0.3cm}
    \caption{The twisted $\textbf{k}$-space. Panel (a) shows the $\mathbf{k}$-grid of a 2D system with $N=26$ free neutrons under PBC (squares), and TBC (circles) with $\boldsymbol{\theta}=2\pi(1,2)/7$. The closed symbols represent the filled momentum states and the closed curve shows the boundary of the filled $\mathbf{k}$-states of the infinite system. Panel (b) shows the number of distinct $\mathbf{k}$-magnitudes generated by different twisted BC.}
\end{figure}

The energy of the infinite system diverges, being an extensive quantity. The relevant intensive quantities are the energy density and the energy per particle which are connected through
\begin{align}
    \frac{E_{\textrm{TL}}}{V} = n \frac{E_{\textrm{TL}}}{N}~.
\end{align}
where the energy density is calculated as the TL limit of Eqs.~(\ref{eq:SwaveEnergyBCS}) or
~(\ref{eq:bSwaveEnergyBCS}):
\begin{equation}
     \frac{E_{\textrm{TL}}}{V} = \frac{1}{2\pi ^2}\int _0^{\infty} dk k^2 2v^2(k) \epsilon (k) + \\ 
        + \frac{1}{\pi ^3} \int _0^{\infty} dk dk' k^2 {k'}^2 V_0(k,k')u(k)v(k)u(k')v(k') \label{eq:EnergyBCS_TL}~.
\end{equation}

It should be noted that, both the BCS and PBCS ground-states converge to the same state at TL \cite{Palkanoglou:2020}. The difference lies in that BCS approaches the infinite system through a grand canonical ensemble while PBCS does so in a canonical ensemble. This means that Eq.~(\ref{eq:EnergyBCS_TL}) could be calculated as the TL limit of Eqs.~(\ref{eq:SwaveEnergyPBCS}) or~(\ref{eq:bSwaveEnergyPBCS}), equivalently.
\subsubsection{The solution of the BCS gap equations}
The gap equations of BCS come from minimizing the free energy of the BCS ground state with respect to the pair occupation distributions. As mentioned above, the normal state corresponds to $\Delta (\mathbf{k}) =0$ and so when solving the gap equations, for a given interaction, one is inquiring about the existence of pairing correlations and gets an answer in the form of the gap distribution: the trivial solution represents the absence of pairing. The gap equations for a finite system with an even number of particles are Eqs.~(\ref{eq:Gap1BCSSwave})~\&~(\ref{eq:Gap2BCSSwave}), for a system with an odd number of particles are Eqs.~(\ref{eq:bGap1BCS})~\&~(\ref{eq:bGap2BCS}), and for an infinite superfluid they turn into Eqs.~(\ref{eq:TL_SwaveGap1BCS})~\&~(\ref{eq:TL_SwaveGap2BCS}). Regardless of the nature of the superfluid system (even, odd, or infinite), the gap equations are a set of non-linear coupled equations and so they have to be solved self-consistently. We will describe only the method used for the even particle-numbers, since it is identical to the ones used for odd particle-numbers and the TL. With the solutions of the equations being $\Delta(\mathbf{k})$ and $\mu$, one can solve the first gap equation, Eq.~(\ref{eq:Gap1BCSSwave}), iteratively for a given  $\mu$ and use the resulting $\Delta(\mathbf{K})$ in the second gap equation, Eq.~(\ref{eq:Gap2BCSSwave}), and calculate $\left<N\right>$, thus reducing the problem to the solution of the equation:
\begin{equation}
\left<N\right>(\mu)=N_0  ~. \label{eq:Solution}  
\end{equation}
 The root of Eq.~(\ref{eq:Solution}) can be found numerically concluding the solution of the BCS gap equations. 
 \begin{figure}[htp]
\begin{center}
\includegraphics[width=0.8\columnwidth,clip=]{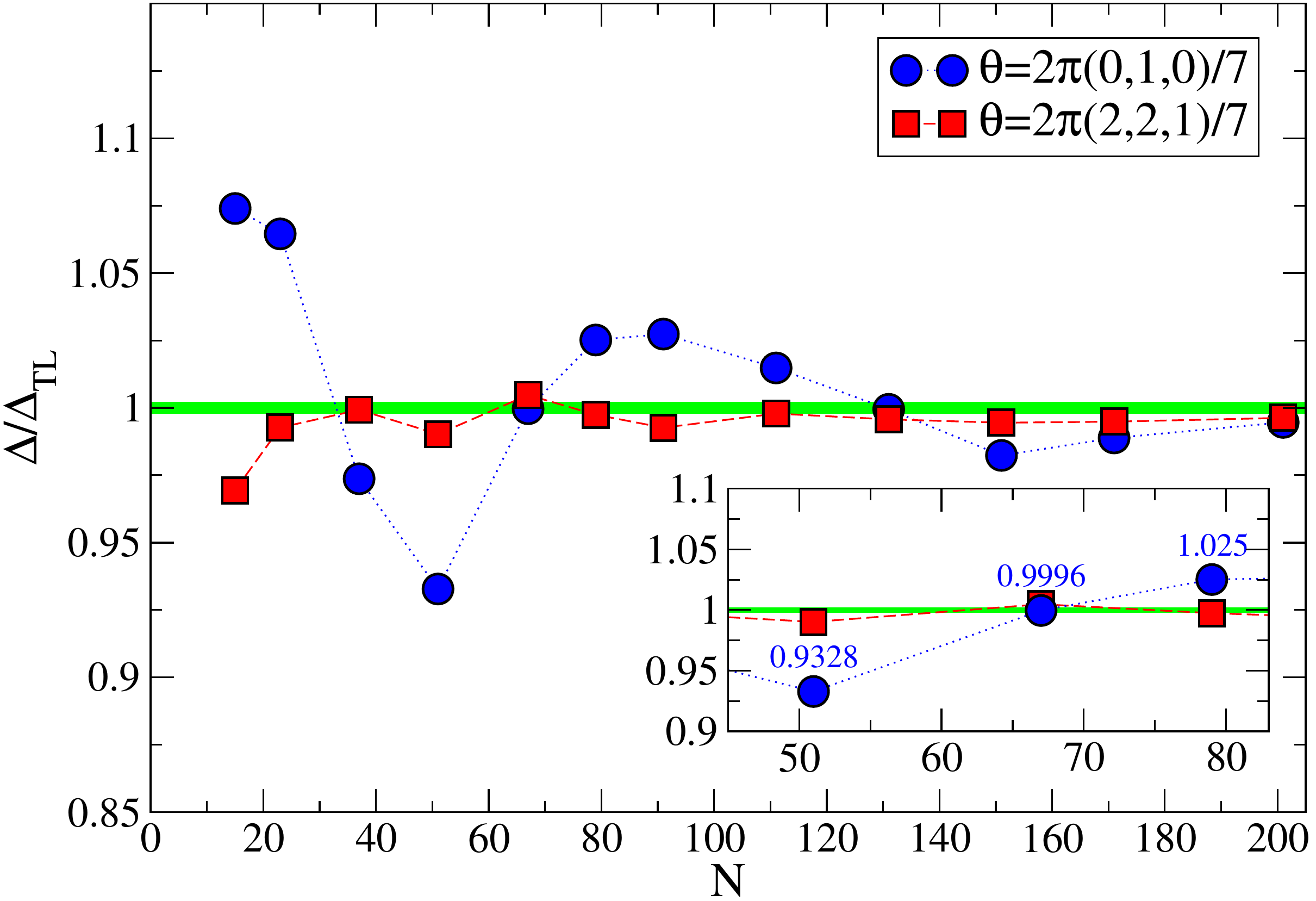}
\caption{The OES for different twist angles $\boldsymbol{\theta}$. The angle $\boldsymbol{\theta}=2\pi (2,2,1)/7$ yields minimum FSE in the OES across all $N$. The angle $\boldsymbol{\theta}=2\pi(0,1,0)/7$ yields minimum FSE for $N=67$, a particle number routinely used in QMC calculations for NM as it has been demonstrated to yield minimum FSE under PBC.	\label{fig:magic_twists}}
\end{center}
\end{figure}

 While it is straightforward to apply identical schemes to the gap equations for odd particle-numbers as well as at the TL, when it comes to blocking an additional step is needed to properly describe the ground state of an odd particle-numbered system. As mentioned above, the blocked state $\mathbf{b}$ that appears in Eqs.~(\ref{eq:bGap1BCS})~\&~(\ref{eq:bGap2BCS}) should be chosen so that it minimizes the energy of the system. For PBCS this energy corresponds to the energy in Eq.~(\ref{eq:bSwaveEnergyPBCS}). This minimization entails the computationally expensive task of solving the blocked gap equations multiple times to acquire the $v_\mathbf{k}$ and $u_\mathbf{k}$ distributions needed for the blocked PBCS energy. An alternative route, suitable for strongly correlated systems, is a perturbative scheme. That is, one can attain an approximation of the pair probability distributions for an odd system by solving the gap equations that correspond to a fully paired system with an odd number of particles. Consequently these distributions can be used repeatedly in Eq.~(\ref{eq:bSwaveEnergyPBCS}) for a range of momentum states $\mathbf{b}$ to identify the state that yields the minimum energy. The error introduced in the distributions $v_\mathbf{k}$ and $u_\mathbf{k}$ through this perturbative approach is inversely proportional to the number of pairs $N_0/2$ and so one can expect that it will not alter the structure of the excitations of the state \cite{book:NuclMBP, Palkanoglou:2020}. With the gap equations solved only once, the use of this perturbative scheme allows for the identification of the proper blocked state in a computationally inexpensive way.


\section{Finite Size Effects and Twisted Boundary Conditions}
\label{sec:FSE}
As a finite system approaches its TL, all intensive quantities of the system approach their TL values and the random trends that they follow are called the Finite-Size Effects (FSE). These effects can be seen in quantities like the energy per particle~\cite{Palkanoglou:2020}, or the pairing gap in Fig.~\ref{fig:D}. One can typically attribute this systematic error to the quantization of $\mathbf{k}$-space induced by the BC: the distributions describing an infinite system are defined on a continuous phase-space and so any approximation that uses distributions defined on a discrete phase-space comes with an error. The origin of one such error can be seen in the FSE observed in the pairing gap in Fig.~\ref{fig:D}, where the curve drawn by the values of $E(\mathbf{k})$ over the grid had a minimum at a momentum which did not coincide with any point on the grid. That resulted in compromising with the closest ``available'' minimum which can jump to the next or the previous $\mathbf{k}$-magnitude with a change in $\left<N\right>$. This is not the only source of FSE in the pairing gap but it is responsible for an additional ``jiggle'' seen in the pairing gap compared to other quantities.
\begin{figure}[htp]
\begin{center}
\includegraphics[width=0.8\columnwidth,clip=]{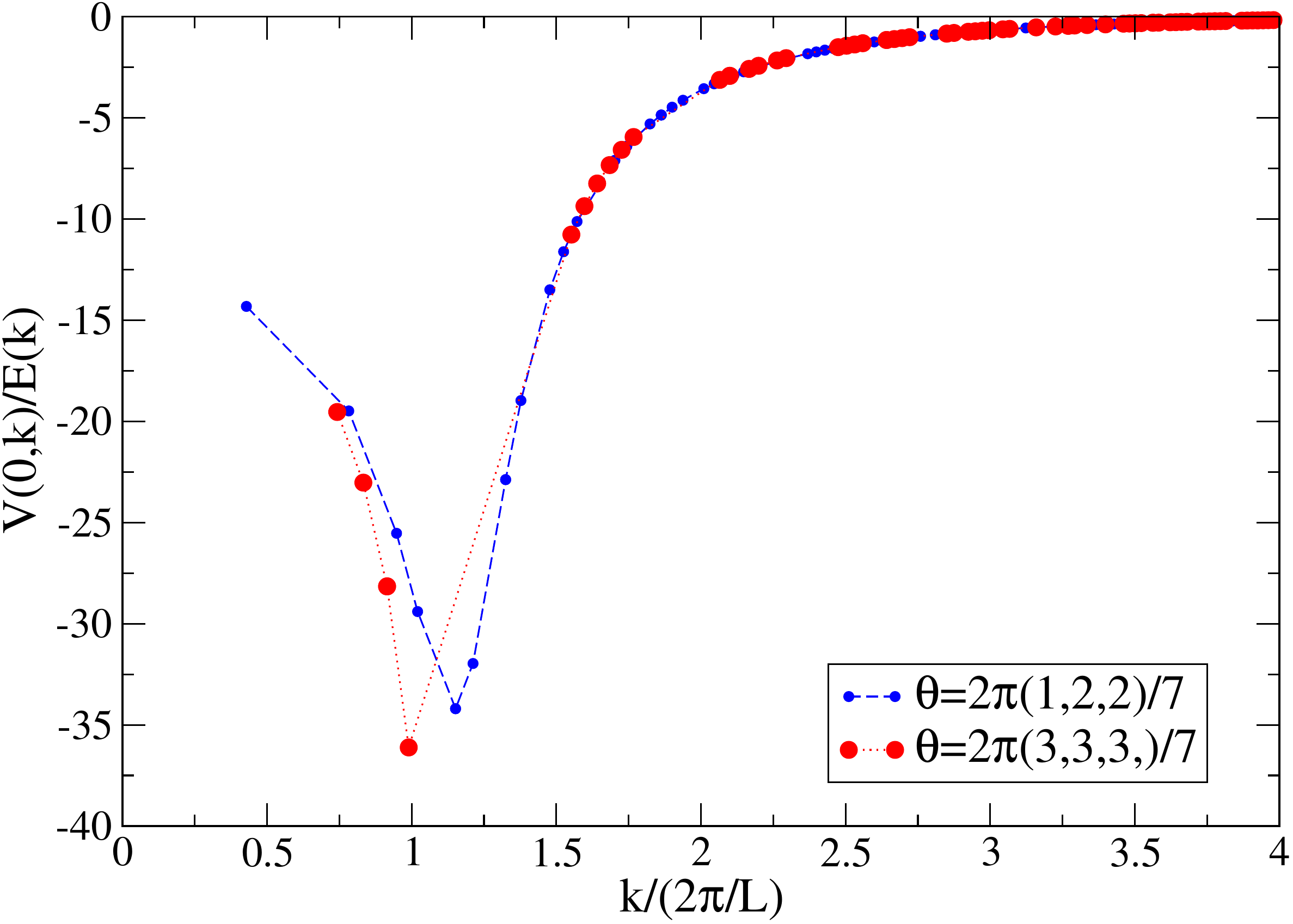}
\caption{The kernel of the gap equation for edge and non-edge cases.	\label{fig:kernel}}
\end{center}
\end{figure}
From the perspective of the OES, these additional FSE have a similar origin: the energy of the odd particle-numbered system is the minimum with respect to the blocked state $\mathbf{b}$ which, in finite systems, increases discontinuously with the particle number $N$. In fact, in systems with odd particle numbers, we observe that the location of the minimum of the quasi-particle energy almost always coincides with the momentum state $\mathbf{b}$ that yields the minimum energy $ E^{\textrm{PBCS}}_{\textrm{odd}}(b;N)$ thus synchronizing the additional FSE seen in the pairing gap and the OES. Quantities like the energy per particle experience less pronounced FSE due to their nature as sums of distributions over the discrete $\mathbf{k}$-space, which makes them less sensitive to small changes in the grid size as the particle number increases.

The systematic errors that are the FSE have to be dealt with when a formulation of the TL is not available and one has to extrapolate from  studies of finite-size systems. In the case of NM, such difficulties are faced by the QMC family of approaches. Quantum Monte Carlo deals with finite systems performing, for the case of NM, computationally expensive simulations whose time complexity scales unfavorably with the particle-number. That allows for tackling systems up to particle numbers of the order of $\sim 100$ and so, when studying infinite matter, one needs to extrapolate to the TL from, oftentimes limited, data of finite systems. This has motivated studies of the FSE using theories which, even though not quantitatively accurate, are qualitatively reliable and can identify a successful extrapolation scheme. For superfluid NM such studies have been carried out both in the BCS~\cite{Gezerlis:2010} and the PBCS~\cite{Palkanoglou:2020} theories and QMC calculations have been conducted using guidance from such studies~\cite{Gezerlis:2008} at great precision. These studies use systems of $N=66$ particles to simulate a system at the TL, an assumption justified by results similar to Fig.~\ref{fig:D} in the context of BCS, and extract the pairing gap using the model-independent OES. With the studies of the FSE rendering them under control for NM, one might wonder whether we can improve the extrapolations employed and actively reduce the observed FSE. The answer lies in the very source of the error: the BC. Indeed, techniques have been developed, initially in the context of solid state physics, that allow for active reduction of the FSE by manipulating the twist $\boldsymbol{\theta}$ introduced in Eq.~(\ref{eq:sp}). In band structure, calculations of properties of infinite periodic solids integrate over the first Brillouin zone, an integration which, in insulators, has been replaced by the choice of a ``special k-point''~\cite{Baldereschi:1973} or a grid of k-points~\cite{Monkhorst:1976}. 
\begin{figure}[htp]
\begin{center}
\includegraphics[width=\columnwidth,clip=]{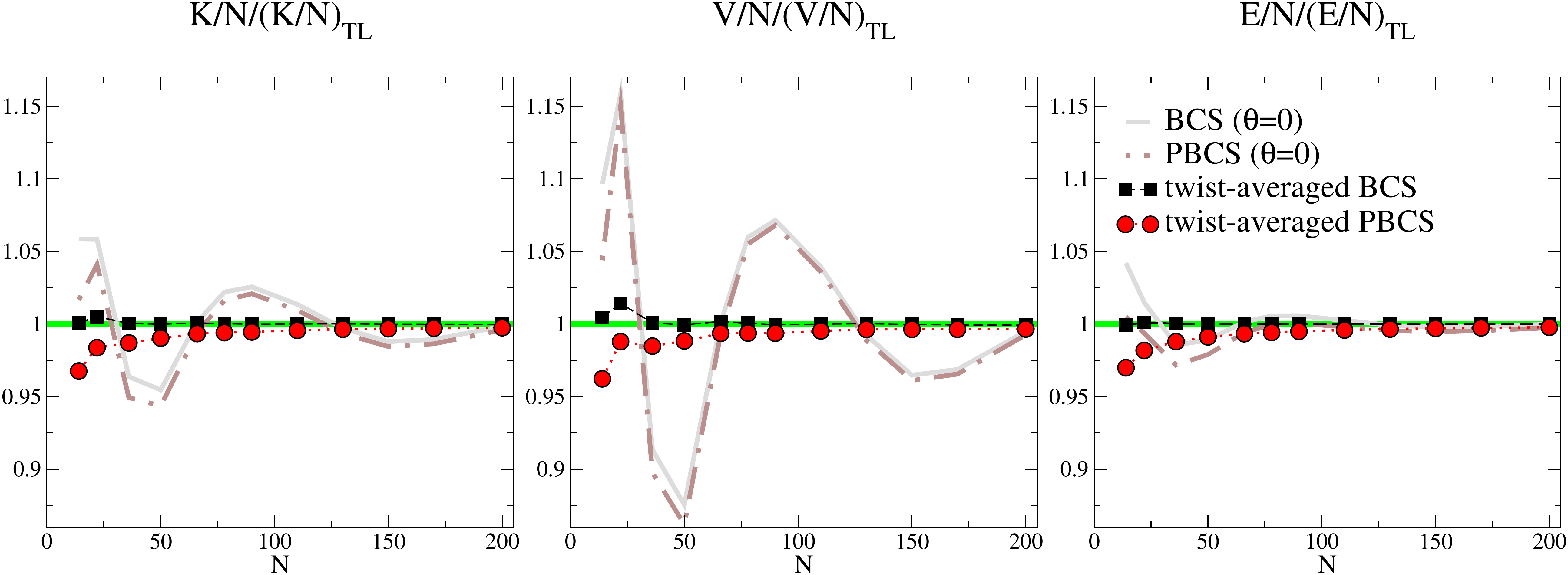}
\caption{The BCS and PBCS energies under PBC and TABC	\label{fig:kve}}
\end{center}
\end{figure}
The integration of the results over the twist is called Twist Averaged Boundary Conditions (TABC) and it has been proven that it will produce exact results in the Hubbard model, in a grand canonical ensemble, also for non-interacting systems~\cite{Gros:1992}, and that it will make certain lattice models converge faster to the TL~\cite{Gammel:1993}. In the context of nuclear matter, TABC has been used for the reduction of FSE in QMC calculations regarding the EoS of nucleonic matter~\cite{Hagen:2014} and in Skyrme-Hartree-Fock calculations of the energy in various nuclear pasta phases~\cite{Schuetrumpf:2016}. To date, there have been no systematic TABC extrapolations for superfluid NM (see, however, Ref.~\cite{Chamel:2010}).

Motivated by their success in other nuclear systems, we applied TBC and TABC in superfluid NM aiming for a reduction of the FSE and to further improve extrapolation schemes. 
In 3D systems under TBC, the twist is a vector with three components each restricted to a $2\pi$ circle. All identical particles are characterized by the same twist, and all quantities are triply periodic~\cite{Byers:1961} in the twist, so that:
\begin{align}
    F(\theta_i + 2\pi) &= F(\theta _i)~, \label{eq:periodic} \\
    -\pi < \theta _i \le \pi~, &\quad i=1,\,2,\,3~. 
\end{align}
When applying the BCS or PBCS theory in the way described in Section~\ref{sec:BCS}, TBC would give rise to a different $\mathbf{k}$-grid through Eq.~(\ref{eq:BC}), i.e., a grid translated by $\theta _i/L$ in the $i$ axis. This is depicted in Fig.~\ref{fig:kspace} where the $\mathbf{k}$-grid of a 2D system of 26 free neutrons has shifted after twisting the BC by $\boldsymbol{\theta} = 2\pi(1,2)/7$. This does not only shift the momentum states available to the different distributions describing the condensate, it also decreases the population $M(k)$ of each $\mathbf{k}$-magnitude. The reason behind this becomes clear if we view the shift of the grid as a shift of the origin in $\mathbf{k}$-space, the reference point of the momentum magnitudes. By shifting the origin one destroys the symmetry once present in $\mathbf{k}$-space making it less likely for momentum states to fall on the same circle with radius $k$. With the momentum cut-off remaining the same, the number of points on the grid is the same and so a smaller population on the different $\mathbf{k}$-magnitudes translates to a larger and denser set of momentum values. This is demonstrated by Fig.~\ref{fig:m_twist} where the ordinal number of the $\mathbf{k}$-magnitude is plotted versus the $\mathbf{k}$-magnitude that it corresponds to, showing the increase in the number of distinct $\mathbf{k}$-magnitudes in TBC. 

Not all twist vectors with components in the $(-\pi,\pi]$ range are independent: the set of twists generated by symmetries of the cubic lattice (i.e., the $\mathbf{k}$-grid) will give rise to identical $\mathbf{k}$-spaces, i.e., identical sets of $\mathbf{k}$-magnitudes and populations. This can be seen directly from Eq.~(\ref{eq:BC}) where, if we let $R$ be a representation of the symmetry group of a cubic lattice, $R$ maps $\mathbf{k}$ to a site of a differently twisted lattice:
\begin{equation}
    R\mathbf{k}_{\boldsymbol{\theta}} = \frac{2\pi}{L} R\mathbf{n} + \frac{R\boldsymbol{\theta}}{L} = \frac{2\pi}{L} \mathbf{n}' + \frac{\boldsymbol{\theta}'}{L} = \mathbf{k}'_{\boldsymbol{\theta}'}~,
\end{equation}
where $\boldsymbol{\theta}'=R\boldsymbol{\theta}$. With the translations being taken care of by restricting $\boldsymbol{\theta}$ in $[0,\pi]$, the only symmetry left to exploit is rotations of $\pi/2$ around each axis. 
\begin{figure}[htp]
\begin{center}
\includegraphics[width=0.8\columnwidth,clip=]{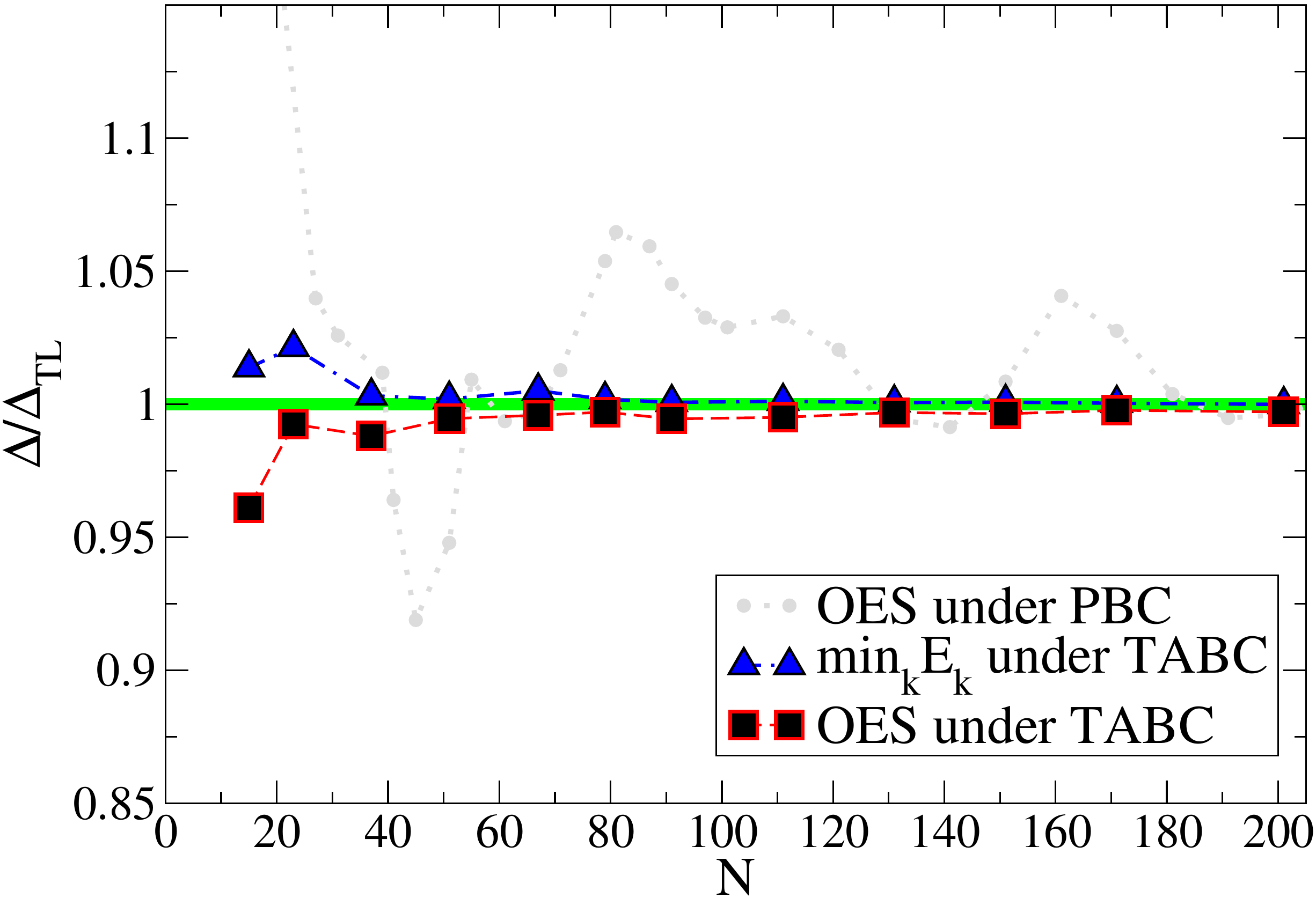}
\caption{The twist-averaged OES for $k_\textrm{F}a=-10$. The twist-averaged minimum of the quasiparticle excitation energy from Fig.~\ref{fig:D} is also shown for comparison.	\label{fig:D_TABC}}
\end{center}
\end{figure}
Therefore, one twist can be taken to represent all eight rotated twists that would generate the same $\mathbf{k}$-space. Such transformations conserve length further justifying why twists connected through them generate momentum spaces that yield the same populations $M(k)$. These arguments can be used to significantly decrease the number of distinct $\boldsymbol{\theta}$-points needed for a twist-integration.

With different TBC giving rise to differently shifted $\mathbf{k}$-spaces, they also generate different FSE since a shifted $\mathbf{k}$-space gives access to different $\mathbf{k}$ magnitudes for the discrete distributions describing the condensate. In principle, as noted above, the resulting $\mathbf{k}$-magnitude structure will be denser and so the FSE for a finite system under TBC are expected to be smaller than those under PBC. Under this light, we performed calculations of various BCS and PBCS quantities under TBC for twist-angles on a grid defined as
\begin{equation}
    \boldsymbol{\theta} = \frac{2\pi}{7}\left(l,m,n\right)~,\quad l,m,n=0,1,\dots ,6~. \label{eq:theta_grid}
\end{equation}
As noted above, twists connected through permutations of $l,m,n$ in Eq.~(\ref{eq:theta_grid}) yield $\mathbf{k}$-spaces connected through a rotation of $\pi/2$ and so they can be considered identical. Thus the values of $l,m,n$ (without an ordering) can uniquely define a set of TBC. For a given vector $\boldsymbol{\theta}$ we have solved the BCS gap equations using the iterative scheme described in sec.~(2.2.4), for a range of particle numbers. The twisted BC shift the momentum grid changing the $\mathbf{k}$-magnitudes that go into Eqs.~(\ref{eq:Gap1BCSSwave})~\&~(\ref{eq:Gap2BCSSwave}) and, consequently, they modify the shell populations $M(k)$. These solutions are then used to obtain quantities like the energy in Eq.~(19).

In the spirit of Ref.~\cite{Baldereschi:1973}, we have identified a special twist-angle that yields minimum FSE in the OES, seen in Fig.~\ref{fig:magic_twists}. This specific twist does not only significantly decrease the overall FSE observed but it also decreases the minimum of FSE seen in the low-$N$ region. These particle-numbers can be used to simulate infinite systems. We have also identified the twist-angle, from the grid in Eq.~(\ref{eq:theta_grid}), that minimizes the FSE at $N=67$. This particle number is commonly used in QMC calculations under PBC~\cite{Gandolfi:2008, Gezerlis:2008} to simulate infinite superfluid NM, owing to its very low FSE~\cite{Palkanoglou:2020, Gezerlis:2008}, $\Delta / \Delta _{\textrm{TL}} \approx 0.997$. We have found that a twist-angle of $\boldsymbol{\theta}=2\pi(0,1,0)/7$ reduces these FSE by $\sim 85\%$ and allows for a more accurate extrapolation to the TL.
\begin{figure}[htp]
\begin{center}
\includegraphics[width=0.8\columnwidth,clip=]{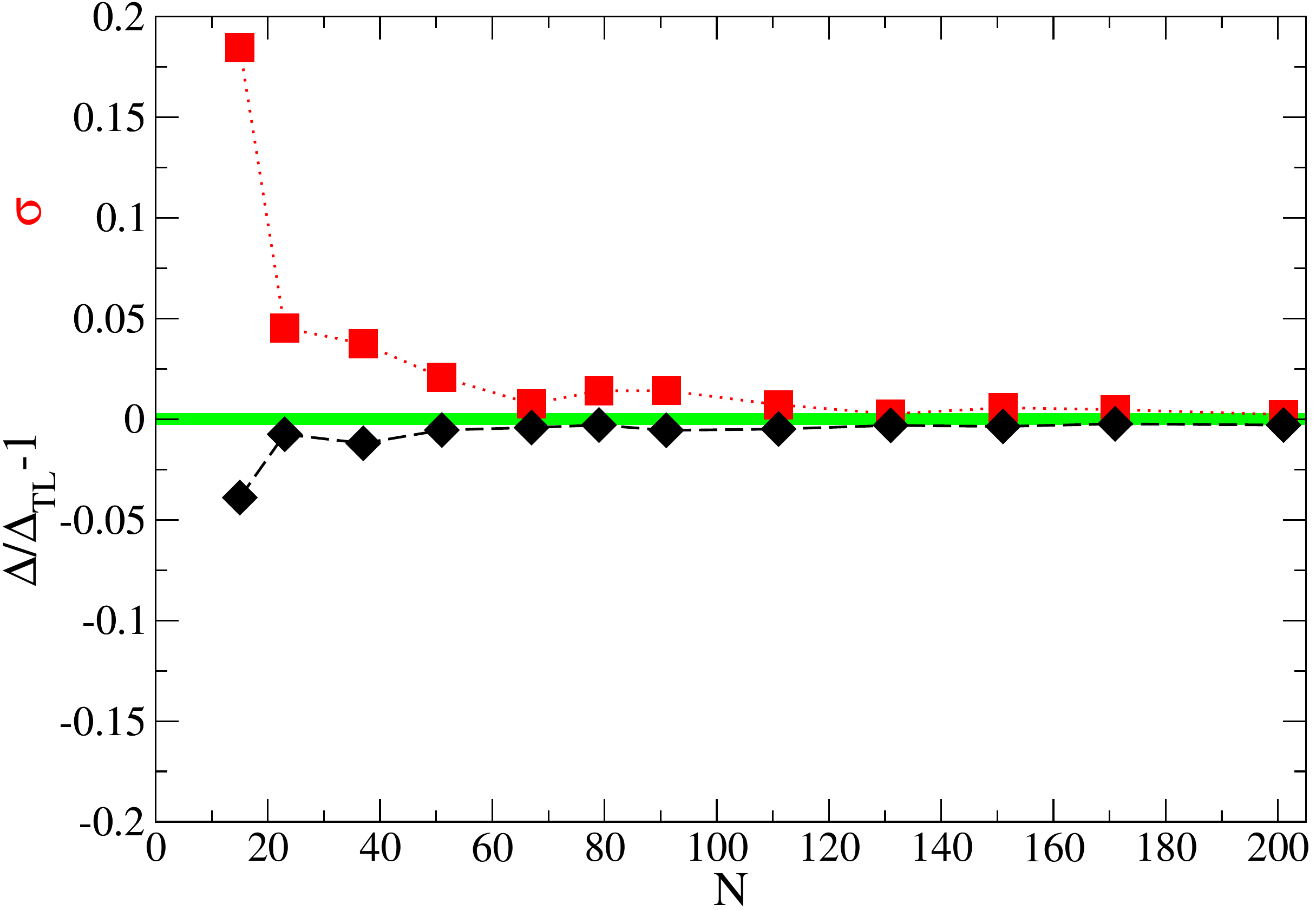}
\caption{The variation $\sigma$ of the OES across the twist-angles compared to the FSE of the twist-averaged OES.	\label{fig:twisted}}
\end{center}
\end{figure}

It is noteworthy that the altered $\mathbf{k}$-magnitude structure created by twisting the BC can give rise to tricky edge cases. For instance, the twists that come closest to the edges of the first Brillouin zone create $\mathbf{k}$-magnitudes with large discontinuities at low-$k$. This proves problematic for the iterative solution of the gap equations, outlined in section~\ref{sec:BCS}. The large discontinuities at low momenta in the gap equations mean that a large part of the non-trivial region of the potential stays unexplored, as seen in Fig.~\ref{fig:kernel} where the kernel of the gap equation in Eq.~(\ref{eq:Gap1BCSSwave}) is plotted for an edge-case together with that of a normal one. Even though this does not impact the solution of the fully-paired gap equation, i.e., Eq.~(\ref{eq:Gap1BCSSwave}) when used for even systems or for odd systems (in the context of a perturbative scheme), it can hinder the convergence of the iterative scheme for the blocked gap equation, Eq.~(\ref{eq:bGap1BCS}). There, blocking the $\mathbf{k}$-state that yields the lowest blocked energy results in decreasing the $\mathbf{k}$-magnitude's population, which is very likely to be 1 for the reasons discussed above, thus removing one more probing point from the non-zero region of the potential, making the situation worse. 

Effectively, this translates to a non-convergence of the iterative scheme for the blocked gap equation for a range of values of the chemical potential wherein the function $\left<N\right>(\mu)$ in Eq.~(\ref{eq:Solution}) becomes ill-defined. In these cases one is forced to use the solution of the perturbative scheme described in section~\ref{sec:BCS}. Even though this introduces a small systematic error, this error can be quantified by comparing perturbative solutions to exact ones in normal cases which indicate a maximum error of $\sim 2.2\%$ in the OES, encountered at low-$N$. With these edge cases making up about $2\%$ of the total number of twist angles considered in the grid of Eq.~(\ref{eq:theta_grid}), this error is irrelevant for approaches like the TABC applied below.

While twisting the BC can significantly reduce FSE, the averaging of the various twist-dependent quantities over a grid of twist-angles, such as the one in Eq.~(\ref{eq:theta_grid}), has been shown to be even more promising in reducing the amplitude of the FSE and flattening the otherwise semi-random trend towards the TL~\cite{Lin:2001}. These are the Twist-Averaged BC where one creates twist-averages of properties of the system:
\begin{equation}
    \left<\hat{F}\right> = \int \frac{d^3\boldsymbol{\theta}}{\left(2\pi\right)^3} \bra{\psi} \hat{F}\ket{\psi} ~. \label{eq:TABC}
\end{equation}
This is done by studying the system under TBC and calculating twist-dependent properties, which are then integrated over the twist angle. Therefore, one can create a twist-averaged OES by calculating the OES under TBC and subsequently integrating Eq.~(\ref{eq:OES}) over the twist space. 

It has been shown~\cite{Lin:2001} that a grand canonical ensemble under TABC produces exact single particle properties for non-interacting systems which is because each momentum state of the infinite system, over the range of the twist angles, occurs precisely once. This guarantees that the non-interacting kinetic energy in the grand canonical ensemble under TABC, is exactly equal to its TL value. With Fermi liquid theory guaranteeing a one-to-one mapping between the low-lying excited states of an interacting system and those of a non-interacting one, we can expect a substantial reduction of FSE for interacting systems in a grand canonical ensemble; this has been argued before in the case of non-superfluid systems~\cite{Gammel:1993, Gros:1992, Lin:2001}. Furthermore, since the projection described in section~\ref{sec:BCS} does commute with the twist-averaging, we anticipate the nice qualities of TABC in a grand canonical ensemble to be present here as well: one can imagine a process where the integration of the twist-averaging, seen in Eq.~(\ref{eq:TABC}), is carried out before the integration of the projection, seen in Eq.~(\ref{eq:ResInt}),  while still employing the grand canonical ensemble. Indeed, as shown in Fig.~\ref{fig:kve}, the BCS energy under TABC has almost no FSE, even at the low-$N$ region. The slightly more intense FSE seen in the BCS potential energy can be traced back to the pronounced FSE which were seen in the minimum of the quasiparticle excitation energy and were discussed in the beginning of this section. That is because the potential energy is proportional to $\sum _{\mathbf{k}\mathbf{l}}F_\mathbf{k}\Delta_{\mathbf{k}\mathbf{l}}$, as can be seen from Eqs.~(\ref{eq:SwaveEnergyBCS})~\&~(\ref{eq:condensation_amplitude}). With the condensation amplitude $F_\mathbf{k}$ peaking at momenta in the region where the minimum of the quasiparticle excitation energy is found, the potential energy is closely related to the pairing gap and as such it inherits some of its FSE. Similar effects are seen in the PBCS energies: the twist-averaging has eliminated almost all of the FSE seen in the calculation under PBC. The remaining difference between the PBCS energy and the TL is the projection-related difference between that and BCS, which under TABC is at the TL for nearly all $N$. This discrepancy is inherent to the symmetry-restoration and it is present even at the non-averaged energies. It cannot be corrected by twist-averaging as clearly seen in Fig.~\ref{fig:R00} where the distribution $\lambda_N^2$ retains its finite spread around $N^*=\left<N\right>$ under TABC, since this spread is proportional to $\sqrt{N^*}$.

Regarding the pairing gap, probed by the model-independent OES, the averaging over the twist-angles achieves substantially reduced FSE, as seen in Fig.~\ref{fig:D_TABC}. Similarly to the energies, this was calculated by numerically integrating over the twist-angle the OES in Eq.~(\ref{eq:OES}) which, under TBC, has a $\boldsymbol{\theta}$ dependence inherited by the energy. The TABC does not only reduce the amplitude of the oscillations observed in the untwisted case, it also brings the gap to its TL value faster. We observe that for $N\gtrapprox 50$ the OES is in a $2\%$ error margin. To achieve similar accuracy over a range of $N$ values under PBC, one has to look at systems well beyond $N=200$. This result is specifically useful to \textit{ab initio} approaches where the study of arbitrarily high particle numbers is not an option. For instance, as noted above, in QMC approaches to NM it is customary to use systems of $N=67$ to extract the TL value of the pairing gap, as per Fig.~\ref{fig:D}. However, under TABC, a slightly better approximation of the TL is provided by $N=79$ neutrons; a system well within reach of modern QMC simulations. This delineates a program of accurately extracting the pairing gap of pure NM in a model-independent fashion via a twist-averaging of the model-independent OES, a quantity accessible by a wide range of \textit{ab initio} techniques. 

The use of TABC has largely eliminated the FSE in the pairing gaps as well. The minimum of the quasiparticle excitation energy under TABC, which is presented in Figs.~\ref{fig:D}~\&~\ref{fig:D_TABC}, is essentially flat compared to the the one calculated under PBC. This is also the case for the OES under TABC presented in Fig.~\ref{fig:D_TABC}. It is noteworthy that, with the twist-averaging illuminating the effects of the projection by reducing the FSE from other sources, the minimum of the quasiparticle excitation energy and the OES differ slightly at low-$N$ under TABC.

The use of the extraction scheme described above in \textit{ab initio} approaches that can handle the complicated correlations present in superfluid NM can provide valuable input for the studies of NS matter. For instance, an \textit{ab initio} study of the static response of strongly paired NM, which has been done before for non-superfluid neutrons \cite{Matt:1, Matt:2}, can use TABC to efficiently eliminate FSE. Furthermore, accurately extracted $^1S_0$ pairing gaps of pure NM can construct effective interactions for energy-density functionals that aim to describe all regions of NSs in a unified way~\cite{Perot:2019, Perot:2020}. Such connections highlight the significance of accurately calculated pure NM pairing gaps.

Finally, in Fig.~\ref{fig:twisted} we present the variance of the set of pairing gaps generated by twisting the BCs according to Eq.~(\ref{eq:theta_grid}):
\begin{equation}
     \sigma^2 = \int \frac{d^3\boldsymbol{\theta}}{\left(2\pi\right)^3} \left[\left<\Delta\right>-\Delta (\boldsymbol{\theta})\right]^2 ~, \label{eq:TABC}
\end{equation}
where the average pairing gap is a twist-average of the kind introduced in Eq.~(\ref{eq:TABC}). Different \textit{ab initio} approaches would be affected differently by the added complexity of the twist-averaging and so one could opt for simple TBC instead of the generally more expensive TABC. Thus, for a given particle number, one can compare the FSE of the twist-averaged pairing gap to the variation of the set of pairing gaps to determine whether a few selected twist-angles could be found and used in an extrapolation scheme instead of a full TABC approach. In other words, Fig.~\ref{fig:twisted} can be used as a guide to inform the choice of TBC or TABC when extracting the pairing gap of infinite NM from a given particle number.

\section{Summary \& Conclusions}
We have performed the first study of superfluid NM under TBC and TABC, techniques used widely in the field of solid state physics, in an effort to prescribe a well-informed extrapolation scheme for \textit{ab initio} approaches that do not have direct access to the infinite system. We applied TABC by averaging calculations of the energy and the pairing gap under TBC, in a range of twist-angles. Our results demonstrate that a twist-averaging approach, compared to PBC, substantially reduces the FSE of the energy and the pairing gap both in amplitude and variation. With the pairing gap being a quantity distinctively sensitive to finite-size, this reduction of the FSE can significantly improve the existent extrapolation schemes in \textit{ab initio} methods, like the QMC family of approaches, where similar studies have dictated systems of $N=67$ neutrons optimal for extrapolation to the TL, under PBC. Given our results, studies with $N=79$ neutrons under TABC would further eliminate the FSE. We also identified special twist-angles which can be used to reduce the FSE under TBC when a TABC study is not within reach. In general, our work provides a more precise extrapolation scheme to the TL for \textit{ab initio} approaches to superfluid NM.

\vspace{6pt} 




\acknowledgments{The authors thank N.~Chamel for putting together this special issue. They would also  like  to  acknowledge  insightful  conversations  with  F. Diakonos and S. Pitsinigkos.  This  work  was  supported  by the Natural Sciences and Engineering Research Council (NSERC) of Canada, the Canada Foundation for Innovation  (CFI),  and  the  Early  Researcher  Award  (ERA) program of the Ontario Ministry of Research, Innovationand Science.  Computational resources were provided by SHARCNET and NERSC.}

\conflictsofinterest{The authors declare no conflict of interest.} 

\abbreviations{The following abbreviations are used in this manuscript:\\

\noindent 
\begin{tabular}{@{}ll}
NS & Neutron Star \\
NN & Neutron-Neutron \\ 
NM & Neutron Matter \\
BCS & Bardeen Cooper Schrieffer \\
OES & Odd-Even Staggering \\
PBCS & Projected BCS \\
FBCS & Full BCS \\
RG & Renormalization Group \\
QMC & Quantum Monte Carlo\\
TL & Thermodynamic Limit\\
PT & P{\"o}schl-Teller \\
CBF & Correlated Basis Functions \\
FSE & Finite-Size Effects \\
BC & Boundary Conditions \\
TBC & Twisted Boundary Conditions \\
TABC & Twist-Averaged Boundary Conditions
\end{tabular}}

\appendixtitles{no} 
\appendix


\reftitle{References}

\end{document}